\documentclass[conference,compsoc]{IEEEtran}
\ifCLASSOPTIONcompsoc
  \usepackage[nocompress]{cite}
\else
  \usepackage{cite}
\fi

\usepackage{graphicx}
\makeatletter
\let\original@makecaption\@makecaption
\long\def\@makecaption#1#2{%
  \begingroup
  \let\footnotesize\normalsize
  \original@makecaption{#1}{#2}%
  \endgroup
}
\makeatother
\ifCLASSOPTIONcompsoc
  \usepackage[caption=false,font=normalsize,labelfont=sf,textfont=sf]{subfig}
\else
  \usepackage[caption=false,font=normalsize]{subfig}
\fi

\usepackage{amsmath}
\usepackage{amssymb}
\usepackage{mathrsfs}

\usepackage{array}
\usepackage{url}
\usepackage{makecell}
\usepackage[ruled,vlined]{algorithm2e}
\usepackage{algpseudocode}
\usepackage{listings}
\usepackage{multirow}
\usepackage{tabularray}
\usepackage[table]{xcolor}
\usepackage{tcolorbox}
\usepackage{fontawesome}
\usepackage{siunitx}
\usepackage{pgfplots}
\usepackage{tikz}
\usetikzlibrary{snakes}
\usepackage{booktabs}
\usepackage{mathtools}
\usepackage{pifont}
\newcommand{\Rmnum}[1]{\uppercase\expandafter{\romannumeral #1}}  
\newcommand{\rmnum}[1]{\romannumeral #1}

\usepackage{booktabs, multirow, makecell, tabularx}
\pgfplotsset{table/search path={data}}
\newlength{\evalfullwidth}
\definecolor{evalblue}{HTML}{2166AC}
\definecolor{evalorange}{HTML}{C65D21}
\definecolor{evalgreen}{HTML}{27836B}
\definecolor{evalpurple}{HTML}{8064A2}
\definecolor{evalgray}{HTML}{B8C2CC}
\pgfplotsset{
  evaluation/.style={
    font=\fontsize{8.5}{10}\selectfont\sffamily,
    label style={font=\fontsize{9}{10}\selectfont\sffamily},
    axis lines*=left, axis line style={black!65,line width=0.5pt},
    tick align=outside, tick pos=left, tick style={black!55,line width=0.4pt},
    major tick length=2.5pt,
    ticklabel style={inner sep=1.5pt},
    scaled ticks=false, /pgf/number format/1000 sep={\,},
    grid style={black!10,line width=0.3pt},
    axis background/.style={fill=white},
    legend style={font=\fontsize{8}{9.5}\selectfont\sffamily,
      draw=none,fill=none,cells={anchor=west},inner sep=1pt},
    legend image post style={line width=0.8pt},
    xlabel near ticks,ylabel near ticks,
    clip marker paths=true,
  },
  evaluation categories/.style={
    evaluation,scale only axis,width=0.64\columnwidth,height=4.8cm,
    ymin=-0.65,ymax=15.65,y dir=reverse,
    ytick={0,...,15},
    yticklabels={FlowLens,Jaqen,nPrintML,NetBeacon,tFusion-RF,Trident-AE,
      MetaTraffic-ARES,ET-BERT,RAPIER,DeepLog,Kitsune,Whisper,
      HyperVision,Undiff,FSC,FAE},
    yticklabel style={font=\fontsize{8}{9.5}\selectfont\sffamily},
    xmajorgrids=true,ymajorgrids=true,
    y grid style={black!7,line width=0.3pt},
    ytick style={draw=none},
    xlabel={Overall accuracy},
  },
  evaluation panel/.style={
    evaluation,scale only axis,width=0.33\columnwidth,height=2.4cm,
    xmajorgrids=false,ymajorgrids=true,
  },
}

\definecolor{takeawayink}{HTML}{234B70}
\definecolor{takeawayfill}{HTML}{EDF3F8}
\newcounter{takeawaycount}
\newsavebox{\takeawaycontent}
\newcommand{\takeaway}[1]{%
  \par\nobreak\addvspace{4pt}\noindent
  \begingroup
  \stepcounter{takeawaycount}%
  \sbox{\takeawaycontent}{%
    \parbox{\dimexpr\linewidth-14pt\relax}{%
      \normalfont\normalsize\raggedright
      {\color{takeawayink}\bfseries Takeaway.} #1\par
      \typeout{TAKEAWAY-LINES \thetakeawaycount: \the\prevgraf}%
    }}%
  \tcbox[on line,
    colback=takeawayfill,colframe=takeawayink,
    boxrule=0.45pt,leftrule=2.5pt,arc=1pt,outer arc=1pt,
    boxsep=0pt,left=6pt,right=5pt,top=3pt,bottom=3pt]{%
    \usebox{\takeawaycontent}}%
  \endgroup
  \par\addvspace{4pt}%
}

\renewcommand{\paragraph}[1]{\smallskip\noindent\textbf{#1.}}
\newcommand{\numberedentry}[2]{%
  \par\addvspace{4pt}%
  \hangindent=1.8em\hangafter=1%
  \noindent\makebox[1.8em][r]{#1)\hspace{0.5em}}%
  \emph{#2}\nobreak\hspace{0.35em}\ignorespaces}

\begin{document}
\title{IDSMorph: Self-Adaptive Malicious Traffic Detection through Consistency-Guided Dynamic Decision Boundaries}
\author{
  \IEEEauthorblockN{
    Hao Fu\IEEEauthorrefmark{1},
    Degang Sun\IEEEauthorrefmark{2},
    Jinxia Wei\IEEEauthorrefmark{2},
    Wei Wan\IEEEauthorrefmark{2},
    Changhai Ou\IEEEauthorrefmark{3},
    Chun Long\IEEEauthorrefmark{2}, and
    Peng Gao\IEEEauthorrefmark{1}
  }
  \IEEEauthorblockA{\IEEEauthorrefmark{1}Virginia Tech, USA\\
  \{haofu26, penggao\}@vt.edu}
  \IEEEauthorblockA{\IEEEauthorrefmark{2}Computer Network Information Center, Chinese Academy of Sciences, China\\
  \{dgsun, weijinxia, wanwei, anquanip\}@cnic.cn}
  \IEEEauthorblockA{\IEEEauthorrefmark{3}Wuhan University, China\\
  ouchanghai@whu.edu.cn}
}

\maketitle

\begin{abstract}
Anomaly-based malicious traffic detection learns decision boundaries for normal behavior and identifies deviations as anomalies. However, normal behavior varies across networks and changes dynamically, limiting adaptability and increasing retraining costs. Despite this variability, our measurements show that clients accessing the same service exhibit consistent flow structures and interaction patterns, since each service provides specific functionality. We therefore construct service-specific decision boundaries. However, variations in personalized requests make it difficult to construct stable boundaries.

To address this challenge, we propose IDSMorph, a self-adaptive malicious traffic detection system that learns consistency-guided embeddings through two key designs: (1) a flow structure consistency builder, which constructs contrastive samples capturing consistent embeddings within services; and (2) a service interaction consistency builder, which captures local structural consistency among different clients and global behavioral consistency among clients accessing that service. IDSMorph constructs dynamic decision boundaries from the majority of similar embeddings within each service and dynamically updates them through a differencing algorithm that mitigates the transient traffic shifts within short observation windows. Experiments show that IDSMorph's detection F1 score decreases by no more than 0.12 points across network environments, compared with an average decrease of 55.43 points for prior methods. In real-world deployment with red-team operators independently choosing attack strategies and timing, IDSMorph achieves an F1 score 46.81 points higher than other methods.

\end{abstract}

\section{Introduction}
\noindent Malicious traffic detection plays a crucial role in network intrusion defense. By learning the decision boundary of normal behavior and flagging traffic beyond this boundary as anomalous, anomaly-based detectors can identify previously unseen attacks~\cite{mirsky2018kitsune,fu2021realtime,fu2023detecting}. 
However, a decision boundary learned from static training data often misclassifies normal traffic as anomalous after deployment, particularly in dynamic environments where user behavior shifts frequently. Keeping decision boundaries aligned with evolving traffic remains an open challenge.

To adapt to benign traffic drift, existing researches broadly fall into two categories.
(i)~\emph{Traffic representation learning} scales up model parameters or adopts complex architectures to capture broader normal flow patterns across diverse applications and protocols~\cite{lin2022bert,zhou2025trafficformer}. Specifically, pre-trained traffic models learn context-aware packet embeddings and intra-flow dependencies from surrounding traffic, yet still rely on fine-tuning for new network environments.
(ii)~\emph{Model adaptation} detects shifts in normal traffic and updates with the newly labeled samples from the deployment network~\cite{han2023owad}. Under continuous traffic evolution, bursty traffic updates the model far more frequently than low-frequency benign access, leading to catastrophic forgetting of historical normal behavior and resulting in false positives. 
In summary, both categories rely on the assumption that training data adequately represents future service access patterns, which rarely holds in real-world deployments.

Since training-enhanced methods struggle to cover dynamic network conditions, we are motivated to construct decision boundaries of normal behavior directly from traffic observed after deployment. Unlike existing approaches that focus on the training phase, we aim to adaptively build decision boundaries tailored to diverse network environments post-deployment, completely avoiding traditional retraining or manual tuning. However, this idea faces a major challenge in that a post-deployment model struggles to build adaptive normal decision boundaries different network behaviors. To address this challenge and construct adaptive dynamic decision boundaries, we aim to identify a stable baseline construction method that remains effective across diverse network environments both before and after deployment. Our measurements reveal that distinct requests to the same service exhibit strong consistency in intra-flow packet distributions and interaction behaviors. For instance, separate Google searches load fixed static assets while maintaining highly similar client-server request-response structures. Therefore, our key insight is to \emph{construct per-service decision boundaries around consistent behaviors} rather than the entire network globally. These consistent behaviors capture stable normal patterns within each service while attack traffic remains discrete and inconsistent, enabling our approach to adaptively build intra-service detection boundaries in dynamic network environments.

\paragraph{Our Work} We present IDSMorph, a self-adaptive malicious traffic detection system based on consistency-guided dynamic decision boundaries. To extract consistent patterns within services across network environments, IDSMorph learns a \emph{flow representation} for shared packet metadata and a \emph{interaction representation} for shared client access behavior. At deployment, IDSMorph constructs and updates a service-specific decision boundary from the resulting embeddings.

The \emph{flow structure consistency builder} aims to extract shared packet patterns across benign accesses to the same service. However, personalized user accesses within a service still produce varying flow patterns that complicate the filtering of user-specific fragments. Instead of examining specific request payloads, we focus on fixed metadata-level variations in packet sequences such as packet loss, duplication, and reordering. Based on these variations, we construct augmented samples and enforce embedding consistency to generalize over potential intra-service traffic variations. To capture fine-grained flow patterns, we further design an encoding model with multiscale feature fusion.

The \emph{service interaction consistency builder} aims to extract consistent behavioral features for services. However, normal access behaviors vary significantly across different services and prevent a unified feature extractor. Instead of learning service-specific behavioral features, we focus on behavioral consistency within each service and across similar services. To capture these relationships, we construct a service interaction graph to represent tripartite server-service-client relationships. Based on this graph, we jointly optimize intra-service consistency embeddings and inter-service consistency embeddings across similar services.

\emph{Dynamic decision boundary construction} aims to adaptively construct detection boundaries for diverse environments. However, normal behaviors vary across services and prevent unified boundary construction. Instead of learning specific normal behavioral features, we leverage majority stable features to establish decision boundaries. Since stable features from a single perspective remain vulnerable, we construct decision boundaries from consistent access behaviors across four perspectives covering global and local behaviors on both server and client that attackers cannot compromise all four boundaries simultaneously. To build reliable boundaries under sparse intra-service observations, we propose a differential algorithm that compares against historical decision boundaries to filter noise and enable adaptive updates.

We evaluate IDSMorph on 14 public traffic datasets against 16 detectors spanning packet representation models \cite{lin2022bert,qing2023low}, flow feature classifiers \cite{holland2021new,zhou2023efficient}, interaction-based detectors \cite{fu2023detecting}, and feature fusion methods \cite{fu2025training,zhao2024trident}. Furthermore, we build a real-world testbed and conduct a month-long deployment to evaluate its practical performance and capability in detecting previously unseen attacks. The results show that:

\par\smallskip\hangindent=1.8em\hangafter=1\noindent\makebox[1.8em][l]{(1)}\textbf{Service-specific boundaries sustain detection accuracy across traffic environments.} By incorporating historical boundary estimates, IDSMorph achieves an overall ROC AUC of 0.9994 on single datasets compared to only 0.9356 for the strongest baseline. Under shifting network conditions across mixed datasets, IDSMorph maintains an overall AUC of 0.9994. Removing dynamic boundary construction causes the largest F1-score drop of 21.42 points.

\par\smallskip\hangindent=1.8em\hangafter=1\noindent\makebox[1.8em][l]{(2)}\textbf{IDSMorph adapts to traffic shifts without target fine-tuning.} Migration across network environments reduces IDSMorph's F1 score by at most 0.12 points, compared with a mean loss of 55.43 points for prior detectors. During continuous observation periods, updating service boundaries limits the F1 decline to one point. Periodic fine-tuning of the baselines yields an overall decline of 33.13 points.

\par\smallskip\hangindent=1.8em\hangafter=1\noindent\makebox[1.8em][l]{(3)}\textbf{IDSMorph detects previously unseen attacks during continuous monitoring.} During our month-long deployment, red-team operators independently determined attack strategies and timing without disclosing attack types or schedules to the defender. IDSMorph detected slow command-and-control beacons and encrypted data exfiltration with an overall deployment F1 of 0.9999. Even after injecting backbone traffic as background noise to increase dynamism, IDSMorph's F1 remaines above 99.04\%, while baseline F1 scores fell below 5\%.

These results show that existing malicious traffic detectors often perform well under idealized laboratory conditions yet struggle in real-world dynamic networks. IDSMorph introduces a novel detection paradigm that constructs decision boundaries directly from the deployment environment by building dynamic baselines from normal access patterns within each individual service upon deployment. By adaptively constructing per-service decision boundaries, IDSMorph enables detection to adapt to diverse dynamic networks without any manual intervention.

\section{Background and Motivation}
\label{sec:problem}

\subsection{Malicious Traffic Detection}
Malicious traffic detection identifies attack-related communications through packet characteristics, flow statistics, and communication relationships \cite{fu2021realtime,fu2023detecting}. Anomaly-based approaches model normal traffic and flag deviations, enabling detection of previously unseen attacks whose behavior differs from the learned reference \cite{mirsky2018kitsune}. However, changes in clients, services, and request patterns can also shift legitimate traffic away from this reference and trigger false alarms. Reliable detection therefore requires a normality reference that tracks evolving benign behavior.

\subsection{Related Work}
\label{app:work}
\paragraph{Feature-Based Methods}
Feature-based detectors summarize packets and flows using header fields, packet lengths, timing, and statistical measurements. These features support efficient classification, including detection on programmable switches \cite{Barradas2021FlowLensEE,liu2021jaqen,zhou2023efficient}. Automated feature representations and cross-modality fusion further combine complementary traffic information for model training \cite{holland2021new,fu2025training}. However, features that distinguish attacks in one environment may lose discriminative power when legitimate applications and traffic composition change.

\paragraph{Sequence and Interaction Modeling}
Sequence-based approaches capture temporal patterns through frequency-domain analysis, recurrent networks, and reconstruction objectives \cite{fu2021realtime,tang2018deep,said2020network}. Autoencoders model dependencies among traffic statistics \cite{mirsky2018kitsune}, while inter-sample comparisons provide additional evidence of anomalous behavior \cite{lian2025facing}. Related work on event sequences uses prediction errors to identify anomalous execution patterns \cite{du2017deeplog}. Graph-based methods extend detection to communication relationships, helping reveal coordinated or multi-stage attacks whose individual flows resemble legitimate activity \cite{fu2023detecting,wang2024agae,do2020apt,nguyen2023new}. However, interaction behavior representations degrade as service and client interaction patterns evolve and invalidate their decision boundaries.

\paragraph{Pretraining and Adaptation}
Pretraining learns reusable traffic representations from large corpora, and downstream adaptation specializes them for individual analysis tasks \cite{lin2022bert,zhou2025trafficformer,cui2025trafficllm}. Other approaches address unknown traffic classes \cite{zhao2024trident}, limited or low-quality training data \cite{he2021deep,qing2023low}, and robustness under changing network conditions \cite{qing2025training}. Drift-aware adaptation detects changes in normal traffic and selects informative samples for labeling and model updates \cite{han2023owad}. However, approaches that rely on target-specific labels or model updates introduce repeated adaptation costss.

\subsection{Threat Model}
\begin{figure}[!t]
    \centering
    \includegraphics[width=\linewidth]{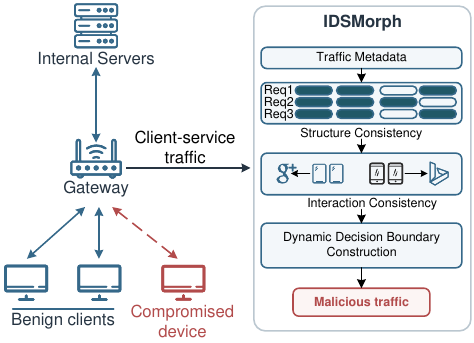} 
    \caption{Deployment setting and threat model. IDSMorph observes client--service traffic at the gateway; attackers control clients that target protected services.}
    \label{fig:threatmodel}
\end{figure}

IDSMorph observes traffic between clients and services at a network gateway (Figure~\ref{fig:threatmodel}). We consider an attacker who knows the services and typical access patterns in the monitored network. From a compromised device, the attacker can target internal servers or communicate with external infrastructure \cite{10.1145/3460120.3484747}. Malicious flows may resemble legitimate service accesses in their packet lengths, timing, and choice of services \cite{fu2023detecting}. We assume that the gateway and detector are trusted, that the attacker controls fewer than half of the devices, and that traffic between uncompromised devices is outside the attacker's control. 

\subsection{Motivation and Challenges}
We measure payload and packet-sequence consistency across service accesses to identify consistent patterns within a service. The measurement covers 10 Google searches for keywords from the 2025 top search list \cite{googletrends}. Individual payloads exhibit high randomness (entropy of approximately 6), and the payload correlation across requests is approximately 0.6 (Figures~\ref{fig:google_data_corr} and \ref{fig:google_entropy_corr}). Figure~\ref{fig:google_access} shows that each visit retrieves static resources such as CSS and JavaScript and follows a similar response structure, with variations in the search results. Figures~\ref{fig:google_pktsize} and \ref{fig:google_pkt_corr} show that packet-size metadata is concentrated around a few values and has a correlation of approximately 0.99 across visits. Network protocols define common communication procedures \cite{rfc9110}, which help explain the consistent flow structures across these service accesses.
\begin{figure}[!t]
    \centering
    \includegraphics[width=\linewidth]{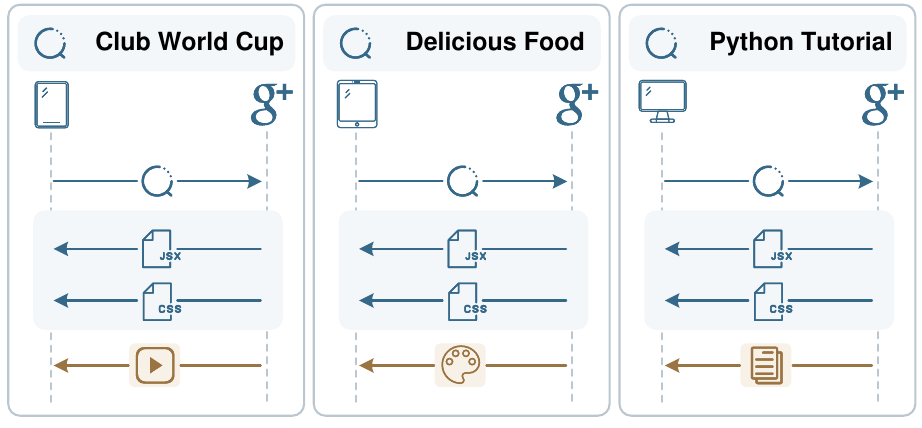}
    \caption{Recurring response structure across Google searches. Requests retrieve similar sequences of static resources despite differences in search results.}
    \label{fig:google_access}
\end{figure}
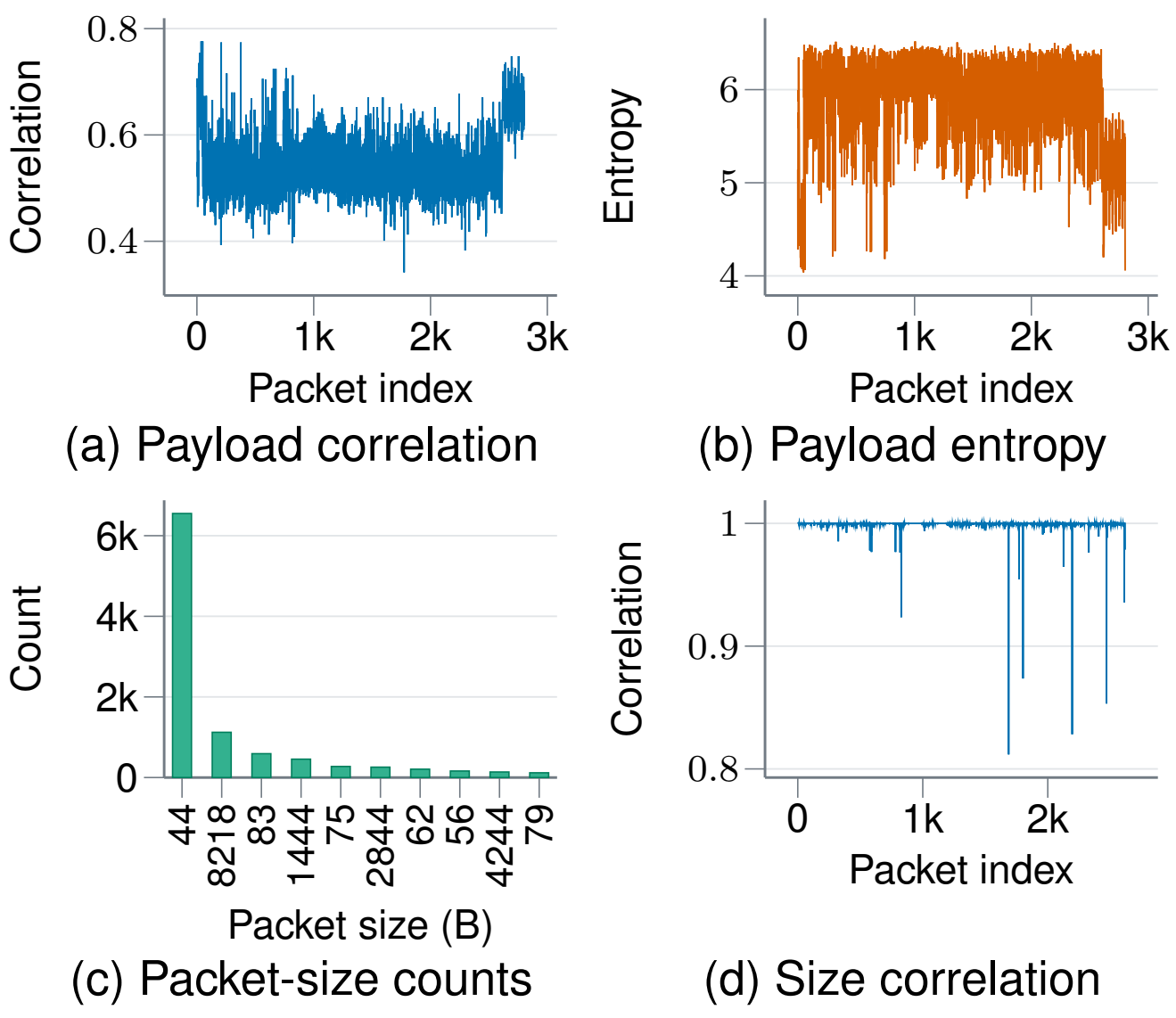
\begin{figure}[!t]
    \centering
    \captionsetup[subfloat]{farskip=2pt,captionskip=2pt}
    \definecolor{googleblue}{HTML}{0072B2}
    \definecolor{googleorange}{HTML}{D55E00}
    \definecolor{googlegreen}{HTML}{009E73}
\definecolor{googleink}{HTML}{263442}
    \definecolor{googlegrid}{HTML}{E2E5E8}
    \definecolor{googlepale}{HTML}{FFFFFF}
    \pgfplotsset{googlepanel/.style={
        font=\fontsize{8.5}{10}\selectfont\sffamily,
        scale only axis, width=0.34\columnwidth, height=0.24\columnwidth,
        axis lines*=left, axis line style={googleink!65, line width=0.55pt},
        axis background/.style={fill=googlepale},
        tick style={googleink!65, line width=0.3pt}, tick align=outside,
        tick pos=left, xmajorgrids=false, ymajorgrids=true, grid style={googlegrid, line width=0.35pt},
        scaled ticks=false, /pgf/number format/1000 sep={\,},
        ylabel near ticks, xlabel near ticks,
        label style={font=\fontsize{9}{10}\selectfont\sffamily, text=googleink},
        ticklabel style={inner sep=1pt},
        xlabel style={yshift=2pt}, ylabel style={yshift=-2pt},
        xtick={0,1000,2000,3000}, xticklabels={0,1k,2k,3k},
        yticklabel style={text width=1.6em, align=right},
    }}
    \subfloat[Payload correlation\label{fig:google_data_corr}]{%
        \makebox[0.48\columnwidth][c]{%
        \begin{tikzpicture}
            \begin{axis}[
                googlepanel,
                xlabel={Packet index},
                ylabel={Correlation},
                ylabel near ticks,
                xlabel near ticks,
            ]
            \addplot[googleblue, line width=0.35pt] table[x expr=\coordindex, y=value, col sep=comma] {google_pktdata_corr.csv};
            \end{axis}
            \pgfresetboundingbox
            \path[use as bounding box]
                ([xshift=-28pt,yshift=-25pt]current axis.south west)
                rectangle ([xshift=3pt,yshift=3pt]current axis.north east);
        \end{tikzpicture}%
        }%
    }
    \hfill
    \subfloat[Payload entropy\label{fig:google_entropy_corr}]{%
        \makebox[0.48\columnwidth][c]{%
        \begin{tikzpicture}
            \begin{axis}[
                googlepanel,
                xlabel={Packet index},
                ylabel={Entropy},
                ylabel near ticks,
                xlabel near ticks,
            ]
            \addplot[googleorange, line width=0.35pt] table[x expr=\coordindex, y=value, col sep=comma] {google_pktentropy_corr.csv};
            \end{axis}
            \pgfresetboundingbox
            \path[use as bounding box]
                ([xshift=-28pt,yshift=-25pt]current axis.south west)
                rectangle ([xshift=3pt,yshift=3pt]current axis.north east);
        \end{tikzpicture}%
        }%
    }

    \subfloat[Packet-size counts\label{fig:google_pktsize}]{%
        \makebox[0.48\columnwidth][c]{%
        \begin{tikzpicture}
            \begin{axis}[
                googlepanel,
                ybar,
                bar width=4pt,
                enlargelimits={0.05,auto},
                xlabel={Packet size (B)}, 
                ylabel={Count},
                ymin=0,
                symbolic x coords={44,8218,83,1444,75,2844,62,56,4244,79},
                xtick=data, xticklabels={44,8218,83,1444,75,2844,62,56,4244,79},
                ytick={0,2000,4000,6000}, yticklabels={0,2k,4k,6k},
                xticklabel style={rotate=90, anchor=east, font=\fontsize{8}{9}\selectfont\sffamily},
                xtick pos=bottom,
                xtick align=outside,
                ylabel near ticks,
                xlabel near ticks,
            ]
            \addplot[fill=googlegreen!80, draw=googlegreen!80!black, line width=0.3pt] coordinates {
                (44, 6551) (8218, 1123) (83, 592)(1444, 456) (75, 275) (2844, 258) (62, 207) (56, 162) (4244, 138)(79, 118)
            };
            \end{axis}
            \pgfresetboundingbox
            \path[use as bounding box]
                ([xshift=-28pt,yshift=-36pt]current axis.south west)
                rectangle ([xshift=3pt,yshift=3pt]current axis.north east);
        \end{tikzpicture}%
        }%
    }
    \hfill
    \subfloat[Size correlation\label{fig:google_pkt_corr}]{%
        \makebox[0.48\columnwidth][c]{%
        \begin{tikzpicture}
            \begin{axis}[
                googlepanel,
                xlabel={Packet index},
                ylabel={Correlation},
                ylabel near ticks,
                xlabel near ticks,
            ]
            \addplot[googleblue, line width=0.35pt] table[x expr=\coordindex, y=value, col sep=comma] {google_pktsize_corr.csv};
            \end{axis}
            \pgfresetboundingbox
            \path[use as bounding box]
                ([xshift=-28pt,yshift=-36pt]current axis.south west)
                rectangle ([xshift=3pt,yshift=3pt]current axis.north east);
        \end{tikzpicture}%
        }%
    }

    \caption{Payload and packet-size characteristics across 10 Google searches: (a) payload correlation, (b) payload entropy, (c) packet-size counts, and (d) packet-size correlation.}
    \label{fig:google_analysis}
\end{figure}

However, using service consistency for malicious traffic detection presents three challenges:

\numberedentry{1}{\textbf{Variations within a Service.}} Payloads vary across requests even when flow structures remain similar (Figures~\ref{fig:google_data_corr} and \ref{fig:google_entropy_corr}). Detection must extract consistent service patterns without overlooking attack differences.

\numberedentry{2}{\textbf{Imitation of Normal Flows.}} Concentrated packet sizes (Figure~\ref{fig:google_pktsize}) provide predictable features that attackers can imitate \cite{fu2023detecting}. Detection must also examine client-service interactions beyond individual-flow features.

\numberedentry{3}{\textbf{Changing Normal Behavior.}} High packet-size correlation still varies across accesses (Figure~\ref{fig:google_pkt_corr}), and a short measurement does not establish long-term stability. Decision boundaries must distinguish transient traffic shifts from lasting changes in service behavior.

\section{IDSMorph Design}
\subsection{System Overview}
\label{section_architecture}
\begin{figure*}[!t]
    \centering
    \includegraphics[width=\linewidth]{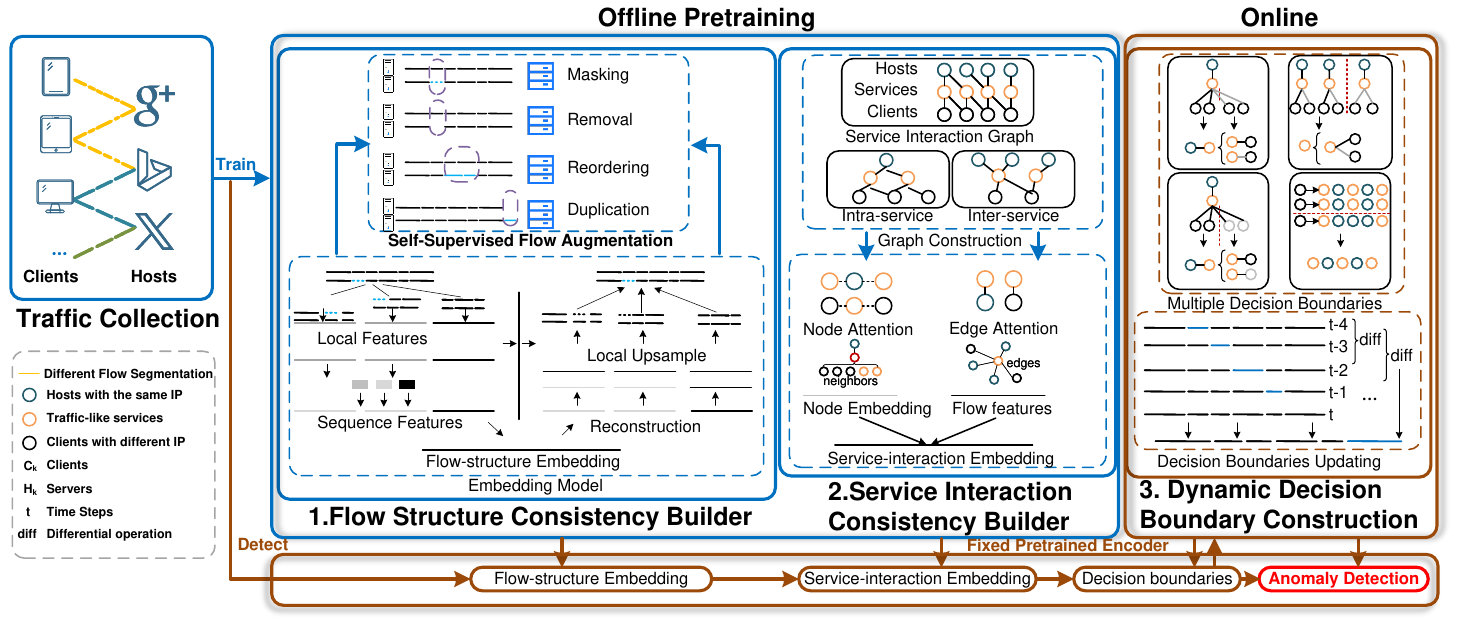} 
    \caption{IDSMorph architecture. Offline pretraining learns flow and interaction representations; online detection constructs and updates decision boundaries from unlabeled traffic while keeping both encoders fixed.}
    \label{fig:sysoverview}
\end{figure*}

To address these challenges, IDSMorph integrates three modules for extracting consistent flow structures, modeling service interactions, and adapting decision boundaries (Figure~\ref{fig:sysoverview}). The embedding modules are pretrained offline and support boundary construction and anomaly detection online.

\paragraph{Flow Structure Consistency Builder} This module extracts consistent flow structures within a service while reducing sensitivity to request variations and transmission perturbations. Section \ref{sec:flow_embedding} describes the encoder and training objectives.

\paragraph{Service Interaction Consistency Builder} This module captures client-service interaction patterns to support detection of attacks that imitate normal flows. A service interaction graph combines communication relationships with flow representations, and graph attention encodes the behavior of connected clients and services. Section \ref{sec:interaction_embedding} describes graph construction, interaction encoding, and contrastive learning.

\paragraph{Dynamic Decision Boundary Construction} This module adapts detection to changes in normal service behavior without retraining the embedding models. Density-based clustering identifies dominant groups of similar embeddings to construct local and global decision boundaries for services and clients. A differencing algorithm uses historical references to mitigate transient traffic shifts during boundary updates. Section~\ref{sec:boundaries} describes boundary construction, updates, and anomaly detection.

\subsection{Flow Structure Consistency Builder}
\label{sec:flow_embedding}
\begin{figure}[!t] 
    \centering 
    \includegraphics[width=\linewidth]{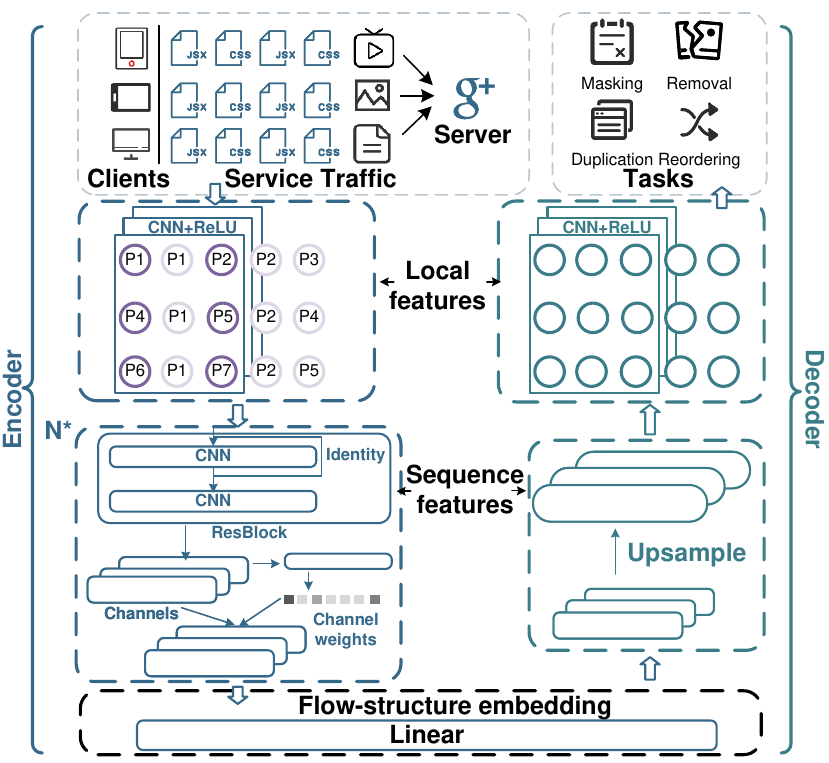} 
    \caption{Flow structure consistency builder and self-supervised training. The encoder maps packet metadata to embeddings; reconstruction, augmented-view consistency, and contrastive objectives guide learning.} 
    \label{fig:au} 
\end{figure}

The flow structure consistency builder learns consistency-guided embeddings of accesses to the same service through pretraining on large-scale unlabeled network traffic while retaining differences between flows. IDSMorph pairs each flow with its perturbed views and minimizes their embedding distance, providing consistency supervision without service labels. Figure~\ref{fig:au} shows the encoder and reconstruction architecture used for this training.

The model input contains packet metadata extracted from raw traffic, excluding payload content and protocol-specific header fields. We parse raw packets $P$ and aggregate them into flows $F$ based on the 5-tuple. A flow of $L$ packets is represented as $F=(\mathbf{x}_1,\ldots,\mathbf{x}_L)$, where $\mathbf{x}_i=(P_{size,i},P_{ts,i},P_{dir,i})$ contains the size, inter-arrival time, and direction of packet $i$. This sequence provides a common input representation across protocols.

\paragraph{Flow Pattern Encoder} To extract consistent flow structures across service accesses, the encoder first captures local packet patterns and then combines them to represent the entire flow (Figure~\ref{fig:au}). Multi-channel convolutions process consecutive packet features $\mathbf{x}_i$ to produce local features $\mathbf{H}^{(0)}$, capturing patterns such as a small outgoing packet followed by several large incoming packets. The same convolutional kernels are applied across packet positions to identify patterns repeated within a flow or shared across accesses.

To capture relationships between local patterns, $N$ residual convolutional blocks progressively transform $\mathbf{H}^{(0)}$ into sequence features $\mathbf{H}^{(N)}$, combining information from more packet positions while retaining earlier local features. Channel attention assigns a learned weight $\alpha_c$ to each feature channel $\mathbf{H}^{(N)}_c$. A linear layer combines the weighted features $\alpha_c\mathbf{H}^{(N)}_c$ into the low-dimensional flow representation $\mathbf{E}_p$, representing both local packet patterns and their relationships across the flow.
The complete encoding process is expressed as
\begin{equation}
\begin{aligned}
    \mathbf{H}^{(N)} &= \operatorname{ResBlocks}_{1:N}\bigl(\operatorname{Conv}(F)\bigr), \\
    \mathbf{E}_p &= \operatorname{Linear}\bigl(\operatorname{vec}(\boldsymbol{\alpha}\odot\mathbf{H}^{(N)})\bigr),
\end{aligned}
\label{equ:flow_encoder}
\end{equation}
where $\operatorname{Conv}$ extracts local features, $\operatorname{ResBlocks}_{1:N}$ denotes the $N$ residual blocks, $\boldsymbol{\alpha}$ contains the learned channel weights, $\odot$ denotes channel-wise weighting, and $\operatorname{vec}$ flattens the weighted features for linear projection.

\paragraph{Flow Reconstruction Decoder} To retain flow-specific sequence information during consistency training, the decoder reconstructs the packet sizes, inter-arrival times, and directions of $F$ from $\mathbf{E}_p$ (Figure~\ref{fig:au}). A projection expands $\mathbf{E}_p$ into a decoding representation $\mathbf{D}$, and a single upsampling module restores features at individual packet positions. Multi-channel convolutions map these features to the reconstructed metadata sequence $\mathbf{E}_F$. The complete decoding process is expressed as
\begin{equation}
\begin{aligned}
    \mathbf{D} &= \operatorname{Proj}_{\mathrm{dec}}(\mathbf{E}_p), \\
    \mathbf{E}_F &= \operatorname{Conv}_{\mathrm{dec}}\bigl(\operatorname{Upsample}(\mathbf{D})\bigr),
\end{aligned}
\label{equ:flow_decoder}
\end{equation}
where $\operatorname{Proj}_{\mathrm{dec}}$ expands the flow representation, $\operatorname{Upsample}$ restores positional features, and $\operatorname{Conv}_{\mathrm{dec}}$ reconstructs the three metadata features at each packet position.

\paragraph{Self-Supervised Flow Augmentation} To accommodate potential changes in user traffic patterns by learning consistent flow structures across request and transmission variations, we generate paired training samples from each unlabeled flow $F$. Four transformations $\mathcal{T}_a$, indexed by $a\in\{A,S,M,D\}$, produce augmented views $F^a$ for masking, reordering, removal, and duplication, respectively (Figure~\ref{fig:au}). Pairing each $F^a$ with $F$ provides observations of the same flow under controlled variations, including for services with few observed accesses.

\numberedentry{1}{Packet Masking ($F^A$).} To reduce dependence on request-specific packet features, $\mathcal{T}_A$ masks randomly selected entries $\mathbf{x}_i$ while retaining their sequence positions. The resulting $F^A$ keeps the original length $L$ but hides part of the metadata. Aligning $F^A$ with $F$ encourages the encoder to derive consistent embeddings from the remaining packet relationships. 

\numberedentry{2}{Packet Reordering ($F^S$).} To reduce sensitivity to local ordering changes caused by concurrency or network congestion, $\mathcal{T}_S$ shuffles packet entries within a local portion of $F$. The reordered sequence $F^S=(\mathbf{x}_{\pi(1)},\ldots,\mathbf{x}_{\pi(L)})$ retains all entries, where $\pi$ changes their order only within the selected portion. Aligning $F^S$ with $F$ trains the encoder to retain broader sequence patterns despite local ordering changes. 

\numberedentry{3}{Packet Removal ($F^M$).} To retain flow structures under packet loss or incomplete observations, $\mathcal{T}_M$ randomly deletes entries from $F$. The shortened view $F^M=(\mathbf{x}_{i_1},\ldots,\mathbf{x}_{i_m})$, with $i_1<\cdots<i_m$ and $m<L$, preserves the order of surviving packets but shifts their positions. Aligning $F^M$ with $F$ reduces dependence on complete packet observations and fixed positions. 

\numberedentry{4}{Packet Duplication ($F^D$).} To reduce sensitivity to repeated packet patterns associated with repeated accesses or retransmissions, $\mathcal{T}_D$ duplicates randomly selected entries $\mathbf{x}_i$, producing an expanded view $F^D$ with more than $L$ entries. Aligning $F^D$ with $F$ reduces the influence of repetition counts on the representation of the underlying flow structure. 

\paragraph{Training Objectives} IDSMorph jointly trains the encoder with three complementary losses. A siamese neural network (SNN \cite{chicco2021siamese}) shares encoder parameters between each flow $F$ and its augmented view $F^a$, producing paired embeddings $\mathbf{E}_p^F$ and $\mathbf{E}_p^{F^a}$.

\numberedentry{1}{Reconstruction Loss ($\mathcal{L}_{\text{rec}}$).} To preserve packet-sequence information, we minimize the masked mean squared error between $F$ and its reconstruction $\mathbf{E}_F$:
\begin{equation}
    \mathcal{L}_{\text{rec}}=
    \frac{\sum_{i=1}^{T}M_i\|\mathbf{x}_i-\mathbf{E}_{F,i}\|_2^2}
    {\sum_{i=1}^{T}M_i},
    \label{equ:flow_reconstruction_loss}
\end{equation}
where $T$ is the padded sequence length, $M_i$ is 1 for observed packets and 0 for padding, and $\mathbf{E}_{F,i}$ is the reconstructed metadata at position $i$.

\numberedentry{2}{Augmentation Consistency Loss ($\mathcal{L}_{\text{tasks}}$).} To reduce sensitivity to the four variations, we combine reconstruction consistency with the embedding distance between $\mathbf{E}_p^F$ and $\mathbf{E}_p^{F^a}$:
\begin{equation}
\begin{aligned}
    \mathcal{L}_{\text{tasks}} = & \sum_{t=1}^{4} \frac{1}{t} \sum_{i=1}^{t} \Big(
    \underbrace{
        \frac{1}{\|\mathbf{M}^{(i)}\|_1} \sum_{t'=1}^{t} M_{t'} \|\mathbf{E}_{F^{t'}} - \mathbf{E}_{F}\|_2^2
    }_{\text{MSE}}
    \Big) \\
    & + \underbrace{
        \sum_{t=1}^{4} \frac{1}{t} \sum_{i=1}^{t} \|\mathbf{E}_p^{F} - \mathbf{E}_p^{F^t}\|_2^2
    }_{\text{SNN}}.
\end{aligned}
\label{equ:task-2}
\end{equation}

\numberedentry{3}{Contrastive Loss ($\mathcal{L}_{\text{nt}}$).} To retain differences between flows, we use NT-Xent \cite{aagren2022nt} with views of the same flow as positive pairs and views of other flows as contrasting samples:
\begin{equation}
    \begin{aligned}
    \mathcal{L}_{\text{nt}}={}&-\frac{1}{|\mathcal{P}_{\mathrm{aug}}|}\\
    &\quad{}\times
    \sum_{(i,j)\in\mathcal{P}_{\mathrm{aug}}}
    \log\frac{\exp\bigl(s(\mathbf{z}_i,\mathbf{z}_j)/\tau\bigr)}
    {\sum_{k\in\mathcal{B}\setminus\{i\}}\exp\bigl(s(\mathbf{z}_i,\mathbf{z}_k)/\tau\bigr)},
    \end{aligned}
    \label{equ:flow_contrastive_loss}
\end{equation}
where $\mathcal{B}$ indexes the views in a batch, $\mathcal{P}_{\mathrm{aug}}$ contains ordered positive pairs, $\mathbf{z}_i$ denotes the flow representation of view $i$, $s$ is cosine similarity, and $\tau$ is the temperature. 

The complete training objective combines the three losses:
\begin{equation}
	\mathcal{L}_{\text{p}}=\mathcal{L}_{\text{rec}}+\mathcal{L}_{\text{tasks}}+\mathcal{L}_{\text{nt}}.
\end{equation}

\subsection{Service Interaction Consistency Builder}
\label{sec:interaction_embedding}
\begin{figure}[ht]
    \centering
    \includegraphics[width=0.85\linewidth]{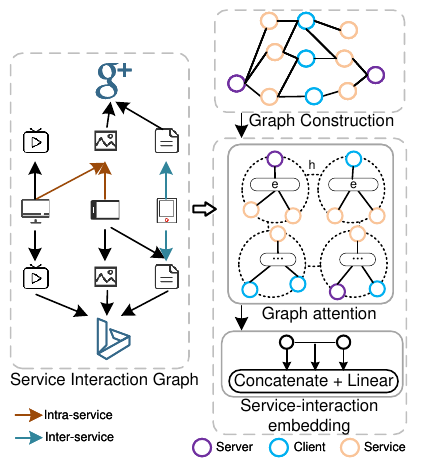}
    \caption{Service interaction consistency builder and contrastive training. Graph attention encodes client--service relationships, with intra-service and inter-service sampling providing consistency supervision.}
    \label{fig:traffic_behavior_embedding}
\end{figure}

The service interaction consistency builder is pretrained on large-scale unlabeled network traffic to capture relationships among service accesses and distinguish flows with similar packet structures but different interaction behavior. Service interaction graphs combine each flow's flow representation with information from its client and service neighbors, while contrastive training learns local structural consistency within services and global behavioral consistency across services (Figure~\ref{fig:traffic_behavior_embedding}).

\paragraph{Service Interaction Graph Construction} To distinguish services hosted at the same IP-port pair, which port-based traffic graphs \cite{fu2025training} group together, IDSMorph clusters flows by access patterns using a density-based method \cite{khan2014dbscan}. Each cluster represents a service instance. To compare packet sequences with different positions, we approximate flow Microelement \cite{fu2025flow} through three complementary measures: \emph{(1) cosine similarity} for overall similarity, \emph{(2) Jaro--Winkler similarity} for local sequence matches, and \emph{(3) a Wasserstein-distance-based measure} for feature distributions. Fusing these scores supports service grouping based on both sequence and distributional similarity.

The resulting graph $\mathcal{G}=(\mathcal{V},\mathcal{E})$ represents clients and service instances as nodes and observed flows as edges. For a flow $F$ from client $u$ to service $v$, its flow representation supplies the edge feature $\mathbf{e}_{uv}=\mathbf{E}_p^F$. Separate service nodes retain differences between accesses to the same server, while shared client nodes connect accesses across services. 

\paragraph{Attention-Based Interaction Encoding} To capture a flow's packet structure and access relationships while emphasizing key traffic features, graph attention selectively aggregates information from each node's neighbors $\mathcal{N}(v)$. At layer $l$, the attention weight $\beta_{uv}^{(l)}$ depends on the endpoint representations $\mathbf{h}_u^{(l)}$, $\mathbf{h}_v^{(l)}$, and the connecting flow feature $\mathbf{e}_{uv}$. All nodes share the same initial representation and update their representations by combining attention-weighted neighbor features to obtain $\mathbf{h}_v^{(l+1)}$. Repeated updates allow a service node to capture behavior across its clients and a client node to capture behavior across its accessed services.

To limit noise propagated through flow edges, we retain the original edge features during node updates. The final client and service embeddings $\mathbf{h}_u$ and $\mathbf{h}_v$ are concatenated with $\mathbf{e}_{uv}$ and projected into the interaction representation $\mathbf{E}_b$:
\begin{equation}
	\mathbf{E}_{b} = \mathcal{W}_{\text{proj}} \left( \mathbf{h}_u \Vert \mathbf{h}_v \Vert \mathbf{e}_{uv} \right),
    \label{equ:grl_2}
\end{equation}
where $\Vert$ denotes concatenation and $\mathcal{W}_{\text{proj}}$ is the learned projection. The resulting embedding retains the flow's packet structure while adding the behavior of its two endpoints, providing interaction context for flows that resemble normal accesses. 

\paragraph{Service Contrastive Learning} To capture consistent interaction patterns among related accesses, we construct intra-service and inter-service pair sets $\mathcal{I}_{\mathrm{intra}}$ and $\mathcal{I}_{\mathrm{inter}}$ from service-specific flow groups $\mathcal{D}_v$, each containing the flows assigned to service node $v$. Service grouping and flow similarity guide pair selection at the two levels:

\numberedentry{1}{Intra-Service Sampling.} We draw pairs $(F_i,F_j)\in\mathcal{D}_v\times\mathcal{D}_v$ to form $\mathcal{I}_{\mathrm{intra}}$, capturing local structural consistency among clients accessing the same service instance.

\numberedentry{2}{Inter-Service Sampling.} We draw pairs $(F_i,F_j)\in\mathcal{D}_v\times\mathcal{D}_{v'}$ to form $\mathcal{I}_{\mathrm{inter}}$, where $v$ and $v'$ represent services with similar access patterns on different server IPs. Pairing accesses to corresponding services on two servers extends global behavioral consistency training beyond a single server address (Figure~\ref{fig:traffic_behavior_embedding}).

To construct both pair sets for services with few observations, we sample with replacement from the relevant groups. Each selected flow is encoded as $\mathbf{E}_b^F$, and the paired embeddings supply the training objectives below.

\paragraph{Training Objectives} Three loss terms train the interaction representations using positive--positive, positive--negative, and negative--negative pair sets, denoted by $\mathcal{P}$, $\mathcal{Q}$, and $\mathcal{R}$. The sampled positive and negative interaction representations are written as $\mathbf{F}_i^+$ and $\mathbf{F}_i^-$, respectively.

\numberedentry{1}{Positive-Pair Consistency.} The squared distance between positive embeddings encourages similar interaction patterns to remain close:
\begin{equation}
\mathcal{L}_{\text{PP}} = \frac{1}{|\mathcal{P}|} \sum_{(i,i') \in \mathcal{P}} \left\| \mathbf{F}_i^+ - \mathbf{F}_{i'}^+ \right\|_2^2.
\end{equation}

\numberedentry{2}{Positive--Negative Comparison.} The positive--negative term compares the similarity of a sampled pair against other negative samples:
\begin{equation}
\mathcal{L}_{\text{PN}} = -\frac{1}{|\mathcal{Q}|} \sum_{(i,j) \in \mathcal{Q}} \log \frac{\exp\big(s(\mathbf{F}_i^+, \mathbf{F}_j^-)/\tau\big)}{\sum_{k} \exp\big(s(\mathbf{F}_i^+, \mathbf{F}_k^-)/\tau\big)}.
\end{equation}

\numberedentry{3}{Negative-Pair Comparison.} The negative--negative term uses negative similarity scores to compare the sampled negative pair with other negative samples:
\begin{equation}
\mathcal{L}_{\text{NN}} = -\frac{1}{|\mathcal{R}|} \sum_{(j,j') \in \mathcal{R}} \log \frac{\exp\big(-s(\mathbf{F}_j^-, \mathbf{F}_{j'}^-)/\tau\big)}{\sum_{k} \exp\big(-s(\mathbf{F}_j^-, \mathbf{F}_k^-)/\tau\big)}.
\end{equation}
Here $s$ and $\tau$ follow the contrastive objective in Section \ref{sec:flow_embedding}. The complete training objective for the service interaction consistency builder is
\begin{equation}
\mathcal{L}_{\text{b}} = \mathcal{L}_{\text{PP}} + \mathcal{L}_{\text{PN}} + \mathcal{L}_{\text{NN}}.
\label{equ:glc_1}
\end{equation}

\subsection{Dynamic Decision Boundary Construction and Update}
\label{sec:boundaries}
\begin{figure}[!t]
    \centering
    \includegraphics[width=0.85\linewidth]{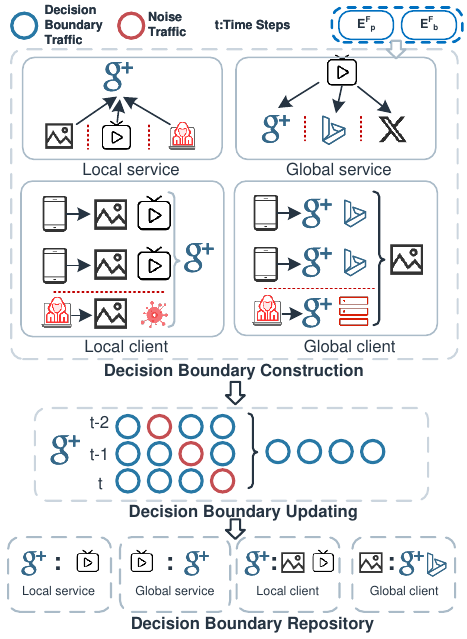}
    \caption{Construction and update of dynamic decision boundaries. Dominant clusters define normal behavior centers for local and global service and client boundaries; historical centers support temporal updates.}
    \label{fig:traffic_baselines}
\end{figure}

The dynamic decision boundary module adapts detection to the target environment by constructing normal behavior centers from the majority of similar embeddings. Each center defines a decision boundary together with a detection threshold. IDSMorph combines local service, global service, local client, and global client boundaries to detect attacks that imitate normal behavior in an individual dimension (Figure~\ref{fig:traffic_baselines}). 

\paragraph{Decision Boundary Construction} To construct boundaries from unlabeled traffic, IDSMorph applies density-based clustering and selects the centroid of the largest cluster as the normal behavior center. Service boundaries are constructed separately using flow representations $\mathbf{E}_p^F$ and interaction representations $\mathbf{E}_b^F$. Client boundaries use sequence features of service accesses within window $t$.

\numberedentry{1}{Local Service Decision Boundary.} To capture normal accesses to service instance $v$, we group the flows observed at $v$ in window $t$ into $\mathcal{D}_{v,t}$. We collect the flow embeddings in $\mathcal{X}_{v,t}^{\mathrm{LS}}=\{\mathbf{E}^F:F\in\mathcal{D}_{v,t}\}$, where $\mathbf{E}^F$ is the flow or interaction representation used for that boundary. Clustering $\mathcal{X}_{v,t}^{\mathrm{LS}}$ produces the center $\mathbf{b}_{v,t}^{\mathrm{LS}}$ of the dominant access pattern within $v$.

\numberedentry{2}{Global Service Decision Boundary.} To identify attacks that imitate or dominate a local service pattern, we group service instances with similar access patterns on different servers into $\mathcal{S}(v)$. Pooling flow embeddings across $\mathcal{S}(v)$ gives $\mathcal{X}_{v,t}^{\mathrm{GS}}=\bigcup_{v'\in\mathcal{S}(v)}\mathcal{X}_{v',t}^{\mathrm{LS}}$. The largest cluster supplies the shared center $\mathbf{b}_{v,t}^{\mathrm{GS}}$, allowing accesses to $v$ to be compared with behavior across related services. This boundary also supports services with sparse local observations, provided that benign accesses dominate the shared cluster.

\numberedentry{3}{Local Client Decision Boundary.} To distinguish clients whose individual flows resemble normal accesses but whose behavior within a service differs, we construct a sequence feature $\mathbf{s}_{u,v,t}$ from client $u$'s accesses to $v$ in window $t$. We collect the client sequence features in $\mathcal{X}_{v,t}^{\mathrm{LC}}=\{\mathbf{s}_{u,v,t}:u\text{ accesses }v\}$. The largest cluster of client sequence features produces $\mathbf{b}_{v,t}^{\mathrm{LC}}$, the center of normal client access behavior within $v$.

\numberedentry{4}{Global Client Decision Boundary.} To capture behavior beyond a single service, we construct a sequence feature $\mathbf{s}_{u,t}$ from each client $u$'s accesses across services in window $t$. Clustering the client features $\mathcal{X}_{t}^{\mathrm{GC}}=\{\mathbf{s}_{u,t}:u\text{ is active in }t\}$ produces $\mathbf{b}_{t}^{\mathrm{GC}}$, the center of normal cross-service client behavior. This dimension supplements local client boundaries when an attacker imitates accesses to an individual service.

Across the four dimensions, the largest cluster $\mathcal{C}_{j,t}$ determines the normal behavior center for boundary $j$:
\begin{equation}
    \widehat{\mathbf{b}}_{j,t} =
    \frac{1}{|\mathcal{C}_{j,t}|}
    \sum_{\mathbf{z}\in\mathcal{C}_{j,t}}\mathbf{z}.
\label{equ:reference_construction}
\end{equation}
The index $j$ identifies a local service, global service, local client, or global client boundary for the corresponding feature group.

\paragraph{Decision Boundary Updating} To mitigate transient traffic shifts caused by transient network conditions and sparse observations, IDSMorph stores historical normal behavior centers $\mathcal{H}_{j,t}=(\mathbf{b}_{j,t-h},\ldots,\mathbf{b}_{j,t-1})$ for each boundary $j$, where $h$ is the history size. An augmented Dickey--Fuller (ADF) test checks the stationarity of $\mathcal{H}_{j,t}$ extended with the newly constructed center $\widehat{\mathbf{b}}_{j,t}$. For a nonstationary series, IDSMorph sets $\mathbf{b}_{j,t}=\widehat{\mathbf{b}}_{j,t}$ to reflect the current traffic. For a stationary series, the temporal model uses $\mathcal{H}_{j,t}$ to estimate $\mathbf{b}_{j,t}$ and mitigate transient traffic shifts. The temporal update is expressed as
\begin{equation}
    B_t = \phi_1 B_{t-1} + \theta_1 \epsilon_{t-1}
          + \theta_2 \epsilon_{t-2} + \epsilon_t,
    \label{equ_arima}
\end{equation}
where $B_t$ is the center value at time $t$, $\epsilon_t$ is the white noise error, and $\phi_1$, $\theta_1$, and $\theta_2$ weight the previous center and error terms. Model parameters are estimated by maximum likelihood. Appendix~\ref{sec:appendix1} examines the history size.

\paragraph{Anomaly Detection}
\label{sec:decision}
IDSMorph flags a flow when any available service or client boundary identifies an anomaly. For each boundary $j$, the corresponding flow embedding or client feature $\mathbf{z}_j(F)$ is compared with its normal behavior center $\mathbf{b}_{j,t}$. The decision rule is
\begin{equation}
    \operatorname{anomaly}(F)=
    \bigvee_{j\in\mathcal{A}_t(F)}
    \left[d_j\bigl(\mathbf{z}_j(F),\mathbf{b}_{j,t}\bigr)>\tau_j\right],
    \label{equ:decision}
\end{equation}
where $\mathcal{A}_t(F)$ contains the available boundaries, $d_j$ is the distance used for boundary $j$, and $\tau_j$ is its threshold. During cold start, detection uses only established boundaries. Flows with $\mathcal{A}_t(F)=\emptyset$ remain unclassified until a boundary becomes available. Appendix~\ref{sec:appendix1} reports threshold sensitivity.

\section{Evaluation}
We evaluate IDSMorph through five research questions. The embedding models remain fixed during online boundary construction and updates. Score differences are reported in points on a 0--100 scale.
\begin{itemize}[\setlength{\IEEElabelindent}{0pt}\setlength{\topsep}{3pt}\setlength{\itemsep}{2pt}\setlength{\parsep}{0pt}\setlength{\partopsep}{0pt}]
    \item \textbf{RQ1. Detection Effectiveness and Evasion Robustness.} How accurately does IDSMorph detect malicious traffic in individual and mixed datasets, including attacks that conceal payloads or imitate normal traffic patterns?
    \item \textbf{RQ2. Transfer and Boundary Adaptation.} Can IDSMorph transfer across network environments and maintain detection as service accesses and traffic composition change, without target-environment fine-tuning?
    \item \textbf{RQ3. Embedding and Boundary Contributions.} How do consistency-guided embeddings and dynamic decision boundaries contribute to detection?
    \item \textbf{RQ4. Pretraining and Detection Costs.} What computational and memory costs does IDSMorph incur during pretraining and detection?
    \item \textbf{RQ5. Detection in an Operational Network.} Can IDSMorph detect red-team attacks amid changing background traffic during a month-long deployment?
\end{itemize}

\subsection{Experimental Setup}
\paragraph{Implementation} We implement IDSMorph in Python 3.12 using dpkt (v1.9.8) for packet parsing and metadata extraction. The embedding model is implemented with PyTorch (v2.9.1), while the graph-based components use PyTorch Geometric (v2.7.0). Clustering modules are built using scikit-learn (v1.7.2). Appendix~\ref{sec:appendix1} reports parameter settings and sensitivity.

\paragraph{Environment} We deploy the IDSMorph prototype on a Dell server equipped with dual Intel(R) Xeon(R) Gold 5218 CPUs and 128 GB DDR4 RAM, running Ubuntu 22.04.5 LTS (Linux v6.8.0). The embedding models are trained and executed on an NVIDIA GeForce RTX 4090 GPU (24 GB VRAM, driver 570.86.10).

\paragraph{Datasets} Our evaluation uses 14 diverse datasets spanning multiple domains and encompassing over a hundred distinct attack types: (\rmnum{1}) DoHBrw-ALL datasets \cite{qing2025training} replay malicious traffic including three types of malicious DoH traffic generated by dns2tcp, DNSCat2, and Iodine. (\rmnum{2}) HyperVision datasets \cite{fu2023detecting}, collected from 92 real-world attack scenarios, including 80 novel malicious traces gathered from over 1500 VPCs in a private cloud environment. (\rmnum{3}) Six additional datasets, including IIoT datasets \cite{ghiasvand2024cicapt} collected from over 40 interconnected devices and more than 15 types of sensors, IoT-DIAD2024 datasets \cite{rabbani2024device} generated from 105 heterogeneous IoT devices, APT2IIoT datasets \cite{ghiasvand2024cicapt} modeling more than 20 unique attack techniques, IDS2018 datasets \cite{sharafaldin2018toward} executed across a simulated enterprise network with 50 attacker machines, IDS2017 datasets \cite{sharafaldin2018toward} conducted over 5 consecutive days using 12 dedicated attacker machines and UNSW-NB15 datasets containing 9 distinct attacks. (\rmnum{4}) USTC-TFC2016 datasets \cite{wang2017malware} composed of 20 distinct traffics. (\rmnum{5}) TON\_IoT datasets \cite{moustafa2021new} integrate telemetry from over 10 heterogeneous IoT sensors deployed across devices running multiple operating systems. (\rmnum{6}) Maple-IDS datasets \cite{li2025malicious} encompasses 7 major categories of cyberattacks targeting N-day vulnerabilities. (\rmnum{7}) Kitsune datasets \cite{mirsky2018kitsune} capture IoT devices cyberattacks. (\rmnum{8}) WEB-IDS23 datasets \cite{lanfer2025technical} spanning 20 distinct attack types, including multiple web application exploits. (\rmnum{9}) NetBeacon datasets \cite{zhou2023efficient} focusing on covert channel attacks and DDoS attacks. IDSMorph's embedding models are pretrained on one day's real-world Internet backbone network by the MAWI project \cite{mawi} in May 2024, which contains 8.1 million flows and 229 billion packets.

\paragraph{Baselines} We compare IDSMorph with and without historical decision boundary references (IDSMorph-h and IDSMorph-nh) against 16 baseline methods. Specifically, content-based methods include ET-BERT \cite{lin2022bert}; interaction-based methods include HyperVision \cite{fu2023detecting}; and anomaly-based methods include Kitsune \cite{mirsky2018kitsune}, Whisper \cite{fu2021realtime}, HyperVision \cite{fu2023detecting}, Undiff \cite{lian2025facing}, FSC \cite{fu2021realtime}, and FAE \cite{fu2021realtime}. We also incorporate hardware-friendly algorithms such as FlowLens \cite{Barradas2021FlowLensEE} and Jaqen \cite{liu2021jaqen}; statistical-feature-based approaches including FlowLens \cite{Barradas2021FlowLensEE}, nPrintML \cite{holland2021new}, and NetBeacon \cite{zhou2023efficient}; and feature fusion methods such as tFusion-RF \cite{fu2025training}, Trident \cite{zhao2024trident}, MetaTraffic-ARES \cite{qing2025training}, RAPIER \cite{qing2023low}, and DeepLog \cite{du2017deeplog}.

\paragraph{Baseline Configuration} All AutoML-based methods \cite{fu2025training,holland2021new} use Random Forest as the underlying model implementation. All deep learning models \cite{lin2022bert,mirsky2018kitsune,lian2025facing,fu2021realtime,zhao2024trident,qing2025training,du2017deeplog}, including IDSMorph, are trained for the same number of epochs and their architectures are carefully tuned to control differences in model size and training budget.

\subsection{RQ1: Detection Effectiveness and Evasion Robustness}
\label{sec:class_eval}
\subsubsection{Detection across Traffic Distributions}
\paragraph{Detection on Individual Datasets} IDSMorph-h achieves an overall AUC of 0.9994, compared with 0.9900 for IDSMorph-nh and 0.9356 for the strongest baseline (Table~\ref{table_evaliation_f1}). These correspond to gains of 6.38 and 5.44 AUC points over that baseline, respectively.

\takeaway{Normal traffic depends on the service. Shared boundaries can mistake benign differences between services for malicious deviations.}

\paragraph{Detection on Mixed Datasets} We augment each dataset with a random 5\% sample from other datasets. The overall AUC remains 0.9994 for IDSMorph-h and decreases from 0.9900 to 0.9887 for IDSMorph-nh, a decrease of 0.13 AUC points (Table~\ref{table_evaliation_f1_2}). The baseline methods show overall AUC decreases of up to 7.72 points, highlighting limited adaptability in realistic mixed-distribution environments.
\begin{table*}[!t]
\centering
\footnotesize
\setlength{\tabcolsep}{2.4pt}
\caption{Detection AUC on individual datasets. Blue and red mark the highest and lowest values in each column. IDSMorph-h and IDSMorph-nh use and omit historical boundary references, respectively.}
\label{table_evaliation_f1}
\renewcommand{\arraystretch}{1.05}
\begin{tabular*}{\textwidth}{@{\extracolsep{\fill}}l|llllllll|llllll|l@{}}
\toprule
\multirow{2}{*}{Model} & \multicolumn{8}{c|}{Multiple attack datasets}                                                     & \multicolumn{6}{c|}{CICDatasets}                                                         & \multirow{2}{*}{Overall} \\
\cmidrule(lr){2-15}
& DoHB & HV & USTC & Kits & TON & WEB & Maple & NetB & IIoT & APT & IDS2018 & IDS2017 & UNSWNB & IOTDIAD &   \\
\midrule
FlowLens               & 0.9562     & 0.7418      & 0.9357       & 0.7725  & 0.9258   & 0.8429    & 0.9635    & 0.8247    & 0.8739   & 0.8639     & 0.7548         & 0.8946       & 0.9658        & 0.8997           &              0.8726            \\
Jaqen                  & \textcolor{red}{0.6358}     & 0.7312      & 0.8425       & 0.8211  & \textcolor{red}{0.7514}   & 0.9085    & 0.7692    & 0.6742    & 0.9382   & 0.7633     & 0.9847         & 0.9914      & 0.8127        & 0.7996           &              0.8160            \\
nPrintML               & 0.9541     & 0.8751      & 0.9124       & 0.9878  & 0.9147   & 0.8975    & 0.9785    & 0.9189    & 0.9384   & 0.7852     & 0.8401         & 0.9878      & 0.9852        & 0.9482           &        0.9231                  \\
NetBeacon              & 0.9587     & 0.8425      & 0.9523       & 0.8797  & 0.8975   & 0.8547    & 0.9257    & 0.8174    & 0.9042   & 0.8076     & 0.9908         & 0.9803      & 0.9788        & 0.9108           &        0.9072                  \\
tFusion-RF             & 0.9627     & 0.9347      & 0.9485       & 0.9751  & 0.9587   & 0.9614    & 0.9452    & 0.8425    & 0.9792   & 0.9128     & 0.8756         & 0.8965      & 0.9382        & 0.9668           &               0.9356           \\
Trident-AE             & 0.8574     & 0.7378      & 0.9565       & 0.9121  & 0.8574   & 0.9268    & 0.9084    & 0.7952    & 0.9287   & 0.9282     & 0.9475         & 0.9634      & 0.8958        & 0.8885           &           0.8931               \\
MetaTraffic-ARES       & 0.8646     & 0.8423      & 0.9874       & 0.9354  & 0.9452   & 0.9174    & 0.8762    & 0.8423    & 0.8527   & 0.8663     & 0.8457         & 0.7532      & 0.8652        & 0.9004           &             0.8782             \\
ET-Bert                & 0.8175     & 0.7425      & 0.9916       & 0.8975  & 0.7529   & \textcolor{red}{0.6125}    & \textcolor{red}{0.7192}     & 0.7412     & 0.8004   & \textcolor{red}{0.6981}     & 0.9875         & 0.9241      & 0.8712        & \textcolor{red}{0.7833}           &              0.8100            \\
RAPIER                 & 0.7028     & 0.8047      & \textcolor{red}{0.6354}       & 0.7474  & 0.9158   & 0.9381    & 0.7428    & 0.7487    & 0.9498   & 0.7582     & 0.8435         & 0.8746      & \textcolor{red}{0.6788}        & 0.7852            &              \textcolor{red}{0.7947}            \\
DeepLog                & 0.8654     & 0.8472      & 0.8756       & 0.9377  & 0.8998   & 0.9458    & 0.7521    & 0.7428    & 0.9004   & 0.8887     & 0.6624         & 0.7624      & 0.8897        & 0.9482           &             0.8513             \\
\midrule
Kitsune                & 0.8752     & \textcolor{red}{0.5143}      & 0.9431       & 0.9041  & 0.8437   & 0.9548    & 0.7862    & \textcolor{red}{0.6074}    & 0.9258   & 0.9306     & 0.6642         & 0.6001      & 0.8752        & 0.9284           &              0.8109            \\
Whisper                & 0.9482     & 0.8152      & 0.8975       & 0.9184  & 0.8736   & 0.9784    & 0.8352    & 0.7784    & \textcolor{red}{0.8287}   & 0.9485     & \textcolor{red}{0.5475}      & 0.6874      & 0.8963        & 0.8472           &              0.8429            \\
Hypervision            & 0.9521     & 0.9394      & 0.8763       & \textcolor{red}{0.6372}  & 0.8971   & 0.9875    & 0.8158    & 0.8857    & 0.9852   & 0.9638     & 0.9745         & 0.9661      & 0.8639        & 0.8175           &              0.8973           \\
Undiff                 & 0.9635     & 0.8651      & 0.9974       & 0.9357  & 0.8425   & 0.9782    & 0.8638    & 0.8047    & 0.9389   & 0.8967     & 0.8742         & 0.8371      & 0.8496        & 0.9585           &              0.9004            \\
FSC                    & 0.8529     & 0.7885      & 0.8427       & 0.7984  & 0.8641   & 0.9687    & 0.8426    & 0.8674    & 0.8759   & 0.9468     & 0.7001         & \textcolor{red}{0.5674}      & 0.9384        & 0.8887           &           0.8388               \\
FAE                    & 0.8625     & 0.7428      & 0.9186       & 0.8754  & 0.8387   & 0.9696    & 0.8593    & 0.7714    & 0.9083   & 0.8889     & 0.6145         & 0.5987      & 0.9185        & 0.9448           &              0.8366            \\
\midrule
IDSMorph-h               & \textcolor{blue}{0.9997}     & \textcolor{blue}{0.9998}      & \textcolor{blue}{0.9999} & \textcolor{blue}{0.9999}  & \textcolor{blue}{0.9999}   & \textcolor{blue}{0.9999}    & \textcolor{blue}{0.9999}    & \textcolor{blue}{0.9999}    & \textcolor{blue}{0.9997}   & \textcolor{blue}{0.9939}     & \textcolor{blue}{1.0000}         & \textcolor{blue}{1.0000}      & \textcolor{blue}{0.9997}        & \textcolor{blue}{0.9998}           &              \textcolor{blue}{0.9994}            \\
IDSMorph-nh              & 0.9985     & 0.9894      & 0.9978       & 0.9478  & 0.9745   & 0.9974    & 0.9968    & 0.9765    & 0.9989   & 0.9914     & 0.9999         & 0.9997      & 0.9954        & 0.9964           &              0.9900            \\
\bottomrule
\end{tabular*}
\end{table*}

\begin{table*}[!t]
\centering
\footnotesize
\setlength{\tabcolsep}{2.4pt}
\caption{Detection AUC on mixed datasets. An asterisk denotes augmentation with a random 5\% sample from other datasets. Blue and red mark the highest and lowest values in each column; -h and -nh denote use and omission of historical boundary references.}
\label{table_evaliation_f1_2}
\renewcommand{\arraystretch}{1.05}
\begin{tabular*}{\textwidth}{@{\extracolsep{\fill}}l|llllllll|llllll|l@{}}
\toprule
\multirow{2}{*}{Model} & \multicolumn{8}{c|}{Multiple attack datasets}                                                     & \multicolumn{6}{c|}{CICDatasets}                                                         & \multirow{2}{*}{Overall} \\
\cmidrule(lr){2-15}
& DoH* & H* & UST* & Kit* & TO* & WE* & Mapl* & Net* & IIo* & AP* & IDS208* & IDS207* & UNSWN* & IOTDIA* &   \\
\midrule
FlowLens               & 0.9439     & 0.7318      & 0.9289       & 0.7642  & 0.9185   & 0.8185    & 0.9582    & 0.8118    & 0.8698   & 0.8529     & 0.7521         & 0.8695       & 0.9489        & 0.8993           &              0.8620            \\
Jaqen                  & \textcolor{red}{0.6085}     & 0.6952      & 0.7691       & 0.7821  & \textcolor{red}{0.6932}   & 0.8109    & \textcolor{red}{0.6692}    & \textcolor{red}{0.5982}    & 0.8609   & \textcolor{red}{0.6392}     & 0.8529         & 0.8368      & 0.8082        & 0.7189           &              \textcolor{red}{0.7388}            \\
nPrintML               & 0.9384     & 0.8129      & 0.9085       & 0.9519  & 0.8695   & 0.8875    & 0.9439    & 0.8692    & 0.9084   & 0.7496     & 0.8096         & 0.9496      & 0.9842        & 0.9158           &        0.8928                  \\
NetBeacon              & 0.9096     & 0.8189      & 0.9289       & 0.8397  & 0.8482   & 0.8159    & 0.9009    & 0.7958    & 0.9085   & 0.8011     & 0.9385         & 0.9728      & 0.9691        & 0.9007           &        0.8820                  \\
tFusion-RF             & 0.9518     & 0.9087      & 0.9351       & 0.9623  & 0.9295   & 0.9059    & 0.9218    & 0.8261    & 0.9638   & 0.8859     & 0.8498         & 0.8682      & 0.9085        & 0.9518           &               0.9121           \\
Trident-AE             & 0.8182     & 0.7348      & 0.9482       & 0.9009  & 0.8319   & 0.9185    & 0.9046    & 0.7869    & 0.9176   & 0.9186     & 0.9287         & 0.9598      & 0.8473        & 0.8692           &           0.8775               \\
MetaTraffic-ARES       & 0.8426     & 0.8089      & 0.9628       & 0.9176  & 0.9152   & 0.9008    & 0.8652    & 0.8162    & 0.8428   & 0.8287     & 0.8389         & 0.7428      & 0.8596        & 0.8763           &             0.8585             \\
ET-Bert                & 0.8078     & 0.7098      & 0.9829       & 0.8879  & 0.7769   & \textcolor{red}{0.6089}   & 0.7079     & 0.7696     & \textcolor{red}{0.7698}   & 0.6782     & 0.9587         & 0.9007      & 0.8696        & 0.7685           &              0.7998            \\
RAPIER                 & 0.6858     & 0.7578      & \textcolor{red}{0.6098}       & 0.7083  & 0.8859   & 0.8692    & 0.7159    & 0.7081    & 0.8529   & 0.7429     & 0.8433         & 0.8459      & \textcolor{red}{0.6508}       & \textcolor{red}{0.7015}            &              0.7556            \\
DeepLog                & 0.8182     & 0.8239      & 0.8598       & 0.9078  & 0.8639   & 0.9089    & 0.7228    & 0.7329    & 0.8875   & 0.8658     & 0.6258         & 0.7289      & 0.8629        & 0.9184           &             0.8234             \\
\midrule
Kitsune                & 0.8529     & \textcolor{red}{0.5008} & 0.9082       & 0.9001  & 0.8319   & 0.9129    & 0.7796    & 0.5928    & 0.9168   & 0.9259     & 0.6138         & 0.5869      & 0.8529        & 0.9007           &              0.7912            \\
Whisper                & 0.8954     & 0.7951      & 0.8496       & 0.8769  & 0.8529   & 0.9085    & 0.8069    & 0.7361    & 0.8084   & 0.9097     & \textcolor{red}{0.5058}         & 0.6529      & 0.8529        & 0.8196           &              0.8051            \\
Hypervision            & 0.8965     & 0.9846      & 0.8185       & \textcolor{red}{0.6089}  & 0.8439   & 0.9105    & 0.8085    & 0.8391    & 0.9152   & 0.8469     & 0.7529         & 0.8605      & 0.8529       & 0.7519           &              0.8351            \\
Undiff                 & 0.9017     & 0.8359      & 0.9259       & 0.9259  & 0.8394   & 0.9629    & 0.8519    & 0.8004    & 0.9182   & 0.8629     & 0.8421         & 0.8219      & 0.8245        & 0.9182           &              0.8737            \\
FSC                    & 0.8182     & 0.7629      & 0.8218       & 0.7789  & 0.8205   & 0.9458    & 0.8329    & 0.8147    & 0.8654   & 0.9839     & 0.6425         & 0.5519      & 0.9263        & 0.8664           &           0.8166               \\
FAE                    & 0.8418     & 0.7109      & 0.8506       & 0.8209  & 0.8159   & 0.9317    & 0.8492    & 0.7529    & 0.8678   & 0.8609     & 0.6081         & \textcolor{red}{0.5236}      & 0.8977        & 0.9304           &              0.8045            \\
\midrule
IDSMorph-h               & \textcolor{blue}{0.9997}     & \textcolor{blue}{0.9998}      & \textcolor{blue}{0.9998}       & \textcolor{blue}{0.9998}  & \textcolor{blue}{0.9999}   & \textcolor{blue}{0.9999}    & \textcolor{blue}{0.9999}    & \textcolor{blue}{0.9999}    & \textcolor{blue}{0.9997}   & \textcolor{blue}{0.9939}     & \textcolor{blue}{0.9999}         & \textcolor{blue}{1.0000}      & \textcolor{blue}{0.9997}        & \textcolor{blue}{0.9998}           &              \textcolor{blue}{0.9994}            \\
IDSMorph-nh              & 0.9954     & 0.9881      & 0.9964       & 0.9471  & 0.9705   & 0.9971    & 0.9968    & 0.9699    & 0.9987   & 0.9908     & 0.9997         & 0.9996      & 0.9954        & 0.9964           &              0.9887            \\
\bottomrule
\end{tabular*}
\end{table*}

\takeaway{Traffic composition can change without changing normal service accesses. Boundary adaptation should distinguish these changes.}

\subsubsection{Detection under Traffic Evasion}
\label{sec:evo_reliable}
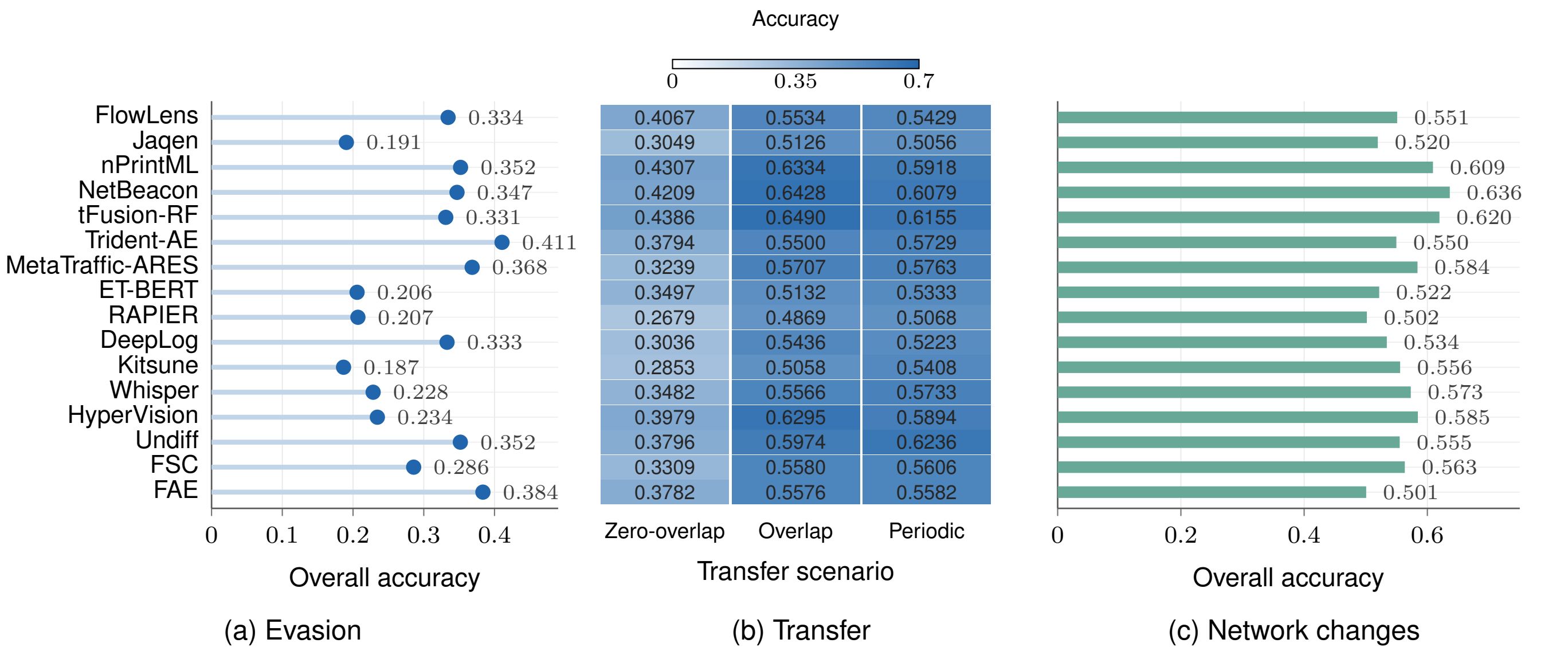
\begin{figure*}[!t]
\centering
\setlength{\evalfullwidth}{\textwidth}
\captionsetup[subfloat]{font=small,farskip=1pt,captionskip=2pt}

\subfloat[Evasion\label{fig:evasion_line}]{\makebox[.36\textwidth][c]{\raisebox{0pt}[170pt][34pt]{\begin{tikzpicture}[baseline=(current axis.south)]
\begin{axis}[evaluation categories,typeset ticklabels with strut,width=0.225\evalfullwidth,height=4.7cm,clip=false,xmin=0,xmax=0.49,xtick={0,0.1,0.2,0.3,0.4},]
\draw[evalblue!28,line width=1.6pt] (axis cs:0,0) -- (axis cs:0.3344,0);
\draw[evalblue!28,line width=1.6pt] (axis cs:0,1) -- (axis cs:0.1906,1);
\draw[evalblue!28,line width=1.6pt] (axis cs:0,2) -- (axis cs:0.3520,2);
\draw[evalblue!28,line width=1.6pt] (axis cs:0,3) -- (axis cs:0.3470,3);
\draw[evalblue!28,line width=1.6pt] (axis cs:0,4) -- (axis cs:0.3311,4);
\draw[evalblue!28,line width=1.6pt] (axis cs:0,5) -- (axis cs:0.4106,5);
\draw[evalblue!28,line width=1.6pt] (axis cs:0,6) -- (axis cs:0.3684,6);
\draw[evalblue!28,line width=1.6pt] (axis cs:0,7) -- (axis cs:0.2057,7);
\draw[evalblue!28,line width=1.6pt] (axis cs:0,8) -- (axis cs:0.2069,8);
\draw[evalblue!28,line width=1.6pt] (axis cs:0,9) -- (axis cs:0.3329,9);
\draw[evalblue!28,line width=1.6pt] (axis cs:0,10) -- (axis cs:0.1866,10);
\draw[evalblue!28,line width=1.6pt] (axis cs:0,11) -- (axis cs:0.2284,11);
\draw[evalblue!28,line width=1.6pt] (axis cs:0,12) -- (axis cs:0.2344,12);
\draw[evalblue!28,line width=1.6pt] (axis cs:0,13) -- (axis cs:0.3517,13);
\draw[evalblue!28,line width=1.6pt] (axis cs:0,14) -- (axis cs:0.2857,14);
\draw[evalblue!28,line width=1.6pt] (axis cs:0,15) -- (axis cs:0.3836,15);
\addplot[only marks,mark=*,mark size=2.3pt,evalblue,nodes near coords={\pgfmathprintnumber[fixed,fixed zerofill,precision=3]{\pgfplotspointmeta}},point meta=x,every node near coord/.append style={font=\fontsize{7.5}{9}\selectfont\sffamily,anchor=west,xshift=3pt,text=black!75}] coordinates {(0.3344,0) (0.1906,1) (0.3520,2) (0.3470,3) (0.3311,4) (0.4106,5) (0.3684,6) (0.2057,7) (0.2069,8) (0.3329,9) (0.1866,10) (0.2284,11) (0.2344,12) (0.3517,13) (0.2857,14) (0.3836,15)};
\end{axis}
\pgfresetboundingbox
\path[use as bounding box] ([xshift=-67pt,yshift=-29pt]current axis.south west) rectangle ([xshift=3pt,yshift=34pt]current axis.north east);
\end{tikzpicture}}}}
\hspace{.01\textwidth}%
\subfloat[Transfer\label{fig:transfer_line_comparison}]{\makebox[.28\textwidth][c]{\raisebox{0pt}[170pt][34pt]{\begin{tikzpicture}[baseline=(current axis.south)]
\begin{axis}[evaluation categories,typeset ticklabels with strut,width=0.255\evalfullwidth,height=4.7cm,
 xmin=-0.5,xmax=2.5,xtick={0,1,2},xticklabels={Zero-overlap,Overlap,Periodic},
 xticklabel style={font=\fontsize{7}{8}\selectfont\sffamily},
 yticklabels={},xlabel={Transfer scenario},xmajorgrids=false,ymajorgrids=false,
 axis lines*=left,axis line style={draw=none},tick style={draw=none},
 colormap={transfer}{color(0cm)=(white);color(1cm)=(evalblue)},
 point meta min=0,point meta max=0.7,
 colorbar horizontal,colorbar style={at={(0.5,1.08)},anchor=south,
 width=0.16\evalfullwidth,height=3pt,xtick={0,0.35,0.7},
 xticklabel style={font=\fontsize{7}{8}\selectfont\sffamily},
 title={Accuracy},title style={font=\fontsize{7}{8}\selectfont\sffamily}}]
\fill[evalblue!58!white] (axis cs:-0.49,-0.49) rectangle (axis cs:0.49,0.49);
\node[font=\fontsize{6.5}{8}\selectfont\sffamily,text=black!85] at (axis cs:0,0) {0.4067};
\fill[evalblue!79!white] (axis cs:0.51,-0.49) rectangle (axis cs:1.49,0.49);
\node[font=\fontsize{6.5}{8}\selectfont\sffamily,text=black!85] at (axis cs:1,0) {0.5534};
\fill[evalblue!78!white] (axis cs:1.51,-0.49) rectangle (axis cs:2.49,0.49);
\node[font=\fontsize{6.5}{8}\selectfont\sffamily,text=black!85] at (axis cs:2,0) {0.5429};
\fill[evalblue!44!white] (axis cs:-0.49,0.51) rectangle (axis cs:0.49,1.49);
\node[font=\fontsize{6.5}{8}\selectfont\sffamily,text=black!85] at (axis cs:0,1) {0.3049};
\fill[evalblue!73!white] (axis cs:0.51,0.51) rectangle (axis cs:1.49,1.49);
\node[font=\fontsize{6.5}{8}\selectfont\sffamily,text=black!85] at (axis cs:1,1) {0.5126};
\fill[evalblue!72!white] (axis cs:1.51,0.51) rectangle (axis cs:2.49,1.49);
\node[font=\fontsize{6.5}{8}\selectfont\sffamily,text=black!85] at (axis cs:2,1) {0.5056};
\fill[evalblue!62!white] (axis cs:-0.49,1.51) rectangle (axis cs:0.49,2.49);
\node[font=\fontsize{6.5}{8}\selectfont\sffamily,text=black!85] at (axis cs:0,2) {0.4307};
\fill[evalblue!90!white] (axis cs:0.51,1.51) rectangle (axis cs:1.49,2.49);
\node[font=\fontsize{6.5}{8}\selectfont\sffamily,text=black!85] at (axis cs:1,2) {0.6334};
\fill[evalblue!85!white] (axis cs:1.51,1.51) rectangle (axis cs:2.49,2.49);
\node[font=\fontsize{6.5}{8}\selectfont\sffamily,text=black!85] at (axis cs:2,2) {0.5918};
\fill[evalblue!60!white] (axis cs:-0.49,2.51) rectangle (axis cs:0.49,3.49);
\node[font=\fontsize{6.5}{8}\selectfont\sffamily,text=black!85] at (axis cs:0,3) {0.4209};
\fill[evalblue!92!white] (axis cs:0.51,2.51) rectangle (axis cs:1.49,3.49);
\node[font=\fontsize{6.5}{8}\selectfont\sffamily,text=black!85] at (axis cs:1,3) {0.6428};
\fill[evalblue!87!white] (axis cs:1.51,2.51) rectangle (axis cs:2.49,3.49);
\node[font=\fontsize{6.5}{8}\selectfont\sffamily,text=black!85] at (axis cs:2,3) {0.6079};
\fill[evalblue!63!white] (axis cs:-0.49,3.51) rectangle (axis cs:0.49,4.49);
\node[font=\fontsize{6.5}{8}\selectfont\sffamily,text=black!85] at (axis cs:0,4) {0.4386};
\fill[evalblue!93!white] (axis cs:0.51,3.51) rectangle (axis cs:1.49,4.49);
\node[font=\fontsize{6.5}{8}\selectfont\sffamily,text=black!85] at (axis cs:1,4) {0.6490};
\fill[evalblue!88!white] (axis cs:1.51,3.51) rectangle (axis cs:2.49,4.49);
\node[font=\fontsize{6.5}{8}\selectfont\sffamily,text=black!85] at (axis cs:2,4) {0.6155};
\fill[evalblue!54!white] (axis cs:-0.49,4.51) rectangle (axis cs:0.49,5.49);
\node[font=\fontsize{6.5}{8}\selectfont\sffamily,text=black!85] at (axis cs:0,5) {0.3794};
\fill[evalblue!79!white] (axis cs:0.51,4.51) rectangle (axis cs:1.49,5.49);
\node[font=\fontsize{6.5}{8}\selectfont\sffamily,text=black!85] at (axis cs:1,5) {0.5500};
\fill[evalblue!82!white] (axis cs:1.51,4.51) rectangle (axis cs:2.49,5.49);
\node[font=\fontsize{6.5}{8}\selectfont\sffamily,text=black!85] at (axis cs:2,5) {0.5729};
\fill[evalblue!46!white] (axis cs:-0.49,5.51) rectangle (axis cs:0.49,6.49);
\node[font=\fontsize{6.5}{8}\selectfont\sffamily,text=black!85] at (axis cs:0,6) {0.3239};
\fill[evalblue!82!white] (axis cs:0.51,5.51) rectangle (axis cs:1.49,6.49);
\node[font=\fontsize{6.5}{8}\selectfont\sffamily,text=black!85] at (axis cs:1,6) {0.5707};
\fill[evalblue!82!white] (axis cs:1.51,5.51) rectangle (axis cs:2.49,6.49);
\node[font=\fontsize{6.5}{8}\selectfont\sffamily,text=black!85] at (axis cs:2,6) {0.5763};
\fill[evalblue!50!white] (axis cs:-0.49,6.51) rectangle (axis cs:0.49,7.49);
\node[font=\fontsize{6.5}{8}\selectfont\sffamily,text=black!85] at (axis cs:0,7) {0.3497};
\fill[evalblue!73!white] (axis cs:0.51,6.51) rectangle (axis cs:1.49,7.49);
\node[font=\fontsize{6.5}{8}\selectfont\sffamily,text=black!85] at (axis cs:1,7) {0.5132};
\fill[evalblue!76!white] (axis cs:1.51,6.51) rectangle (axis cs:2.49,7.49);
\node[font=\fontsize{6.5}{8}\selectfont\sffamily,text=black!85] at (axis cs:2,7) {0.5333};
\fill[evalblue!38!white] (axis cs:-0.49,7.51) rectangle (axis cs:0.49,8.49);
\node[font=\fontsize{6.5}{8}\selectfont\sffamily,text=black!85] at (axis cs:0,8) {0.2679};
\fill[evalblue!70!white] (axis cs:0.51,7.51) rectangle (axis cs:1.49,8.49);
\node[font=\fontsize{6.5}{8}\selectfont\sffamily,text=black!85] at (axis cs:1,8) {0.4869};
\fill[evalblue!72!white] (axis cs:1.51,7.51) rectangle (axis cs:2.49,8.49);
\node[font=\fontsize{6.5}{8}\selectfont\sffamily,text=black!85] at (axis cs:2,8) {0.5068};
\fill[evalblue!43!white] (axis cs:-0.49,8.51) rectangle (axis cs:0.49,9.49);
\node[font=\fontsize{6.5}{8}\selectfont\sffamily,text=black!85] at (axis cs:0,9) {0.3036};
\fill[evalblue!78!white] (axis cs:0.51,8.51) rectangle (axis cs:1.49,9.49);
\node[font=\fontsize{6.5}{8}\selectfont\sffamily,text=black!85] at (axis cs:1,9) {0.5436};
\fill[evalblue!75!white] (axis cs:1.51,8.51) rectangle (axis cs:2.49,9.49);
\node[font=\fontsize{6.5}{8}\selectfont\sffamily,text=black!85] at (axis cs:2,9) {0.5223};
\fill[evalblue!41!white] (axis cs:-0.49,9.51) rectangle (axis cs:0.49,10.49);
\node[font=\fontsize{6.5}{8}\selectfont\sffamily,text=black!85] at (axis cs:0,10) {0.2853};
\fill[evalblue!72!white] (axis cs:0.51,9.51) rectangle (axis cs:1.49,10.49);
\node[font=\fontsize{6.5}{8}\selectfont\sffamily,text=black!85] at (axis cs:1,10) {0.5058};
\fill[evalblue!77!white] (axis cs:1.51,9.51) rectangle (axis cs:2.49,10.49);
\node[font=\fontsize{6.5}{8}\selectfont\sffamily,text=black!85] at (axis cs:2,10) {0.5408};
\fill[evalblue!50!white] (axis cs:-0.49,10.51) rectangle (axis cs:0.49,11.49);
\node[font=\fontsize{6.5}{8}\selectfont\sffamily,text=black!85] at (axis cs:0,11) {0.3482};
\fill[evalblue!80!white] (axis cs:0.51,10.51) rectangle (axis cs:1.49,11.49);
\node[font=\fontsize{6.5}{8}\selectfont\sffamily,text=black!85] at (axis cs:1,11) {0.5566};
\fill[evalblue!82!white] (axis cs:1.51,10.51) rectangle (axis cs:2.49,11.49);
\node[font=\fontsize{6.5}{8}\selectfont\sffamily,text=black!85] at (axis cs:2,11) {0.5733};
\fill[evalblue!57!white] (axis cs:-0.49,11.51) rectangle (axis cs:0.49,12.49);
\node[font=\fontsize{6.5}{8}\selectfont\sffamily,text=black!85] at (axis cs:0,12) {0.3979};
\fill[evalblue!90!white] (axis cs:0.51,11.51) rectangle (axis cs:1.49,12.49);
\node[font=\fontsize{6.5}{8}\selectfont\sffamily,text=black!85] at (axis cs:1,12) {0.6295};
\fill[evalblue!84!white] (axis cs:1.51,11.51) rectangle (axis cs:2.49,12.49);
\node[font=\fontsize{6.5}{8}\selectfont\sffamily,text=black!85] at (axis cs:2,12) {0.5894};
\fill[evalblue!54!white] (axis cs:-0.49,12.51) rectangle (axis cs:0.49,13.49);
\node[font=\fontsize{6.5}{8}\selectfont\sffamily,text=black!85] at (axis cs:0,13) {0.3796};
\fill[evalblue!85!white] (axis cs:0.51,12.51) rectangle (axis cs:1.49,13.49);
\node[font=\fontsize{6.5}{8}\selectfont\sffamily,text=black!85] at (axis cs:1,13) {0.5974};
\fill[evalblue!89!white] (axis cs:1.51,12.51) rectangle (axis cs:2.49,13.49);
\node[font=\fontsize{6.5}{8}\selectfont\sffamily,text=black!85] at (axis cs:2,13) {0.6236};
\fill[evalblue!47!white] (axis cs:-0.49,13.51) rectangle (axis cs:0.49,14.49);
\node[font=\fontsize{6.5}{8}\selectfont\sffamily,text=black!85] at (axis cs:0,14) {0.3309};
\fill[evalblue!80!white] (axis cs:0.51,13.51) rectangle (axis cs:1.49,14.49);
\node[font=\fontsize{6.5}{8}\selectfont\sffamily,text=black!85] at (axis cs:1,14) {0.5580};
\fill[evalblue!80!white] (axis cs:1.51,13.51) rectangle (axis cs:2.49,14.49);
\node[font=\fontsize{6.5}{8}\selectfont\sffamily,text=black!85] at (axis cs:2,14) {0.5606};
\fill[evalblue!54!white] (axis cs:-0.49,14.51) rectangle (axis cs:0.49,15.49);
\node[font=\fontsize{6.5}{8}\selectfont\sffamily,text=black!85] at (axis cs:0,15) {0.3782};
\fill[evalblue!80!white] (axis cs:0.51,14.51) rectangle (axis cs:1.49,15.49);
\node[font=\fontsize{6.5}{8}\selectfont\sffamily,text=black!85] at (axis cs:1,15) {0.5576};
\fill[evalblue!80!white] (axis cs:1.51,14.51) rectangle (axis cs:2.49,15.49);
\node[font=\fontsize{6.5}{8}\selectfont\sffamily,text=black!85] at (axis cs:2,15) {0.5582};
\addplot[only marks,mark=*,opacity=0,point meta=explicit] table[x=scenario,y=method,meta=accuracy] {transfer_heatmap.dat};
\end{axis}
\pgfresetboundingbox
\path[use as bounding box] ([xshift=-3pt,yshift=-29pt]current axis.south west) rectangle ([xshift=3pt,yshift=34pt]current axis.north east);
\end{tikzpicture}}}}
\hspace{.01\textwidth}%
\subfloat[Network changes\label{fig:service_baseline_horizontal}]{\makebox[.34\textwidth][c]{\raisebox{0pt}[170pt][34pt]{\begin{tikzpicture}[baseline=(current axis.south)]
\begin{axis}[evaluation categories,typeset ticklabels with strut,width=0.30\evalfullwidth,height=4.7cm,clip=false,yticklabels={},xmin=0,xmax=0.75,xtick={0,0.2,0.4,0.6},]
\addplot[xbar,bar width=3.8pt,fill=evalgreen!70,draw=none,nodes near coords={\pgfmathprintnumber[fixed,fixed zerofill,precision=3]{\pgfplotspointmeta}},point meta=x,every node near coord/.append style={font=\fontsize{7.5}{9}\selectfont\sffamily,anchor=west,xshift=2pt,text=black!75}] coordinates {(0.5510,0) (0.5196,1) (0.6091,2) (0.6364,3) (0.6196,4) (0.5497,5) (0.5840,6) (0.5218,7) (0.5018,8) (0.5342,9) (0.5558,10) (0.5731,11) (0.5846,12) (0.5551,13) (0.5634,14) (0.5007,15)};
\end{axis}
\pgfresetboundingbox
\path[use as bounding box] ([xshift=-3pt,yshift=-29pt]current axis.south west) rectangle ([xshift=3pt,yshift=34pt]current axis.north east);
\end{tikzpicture}}}}
\caption{Baseline detection accuracy under (a) traffic evasion, (b) three transfer scenarios, and (c) network changes. All panels use the method order in (a); heatmap cells show the original accuracy values.}
\label{fig:robustness_combined}
\end{figure*}

\paragraph{Evasion Attack Evaluation} The evasion evaluation tests IDSMorph and the baseline methods against three strategies that conceal malicious traffic: (1) \emph{Proxy encryption}: attackers tunnel malicious traffic through multiple encrypted proxy layers to conceal payload content; (2) \emph{Traffic normalization}: each attack flow is transformed into randomized packet sizes and transmitted at a low rate with jitter to mimic background noise; (3) \emph{Benign behavior imitation}: attackers craft malicious traffic to match the packet length and byte distribution of normal traffic. Baseline methods suffer severe performance degradation when facing adaptive attacks, with F1 scores dropping by more than 66.29 points under the evaluated attacks, whereas IDSMorph's performance degradation is less than 2 points (Figure~\ref{fig:evasion_line}). This resilience stems from two key properties: (\rmnum{1}) its service-specific dynamic baselines are constructed from dominant benign behaviors, limiting the influence of sparse evasion attempts when the benign-reference assumption holds; (\rmnum{2}) its detection relies on interaction representations that capture relationships among service accesses, adding information beyond packet-level statistics. The results concern the evaluated evasion strategies. Section \ref{sec:discussion} discusses the limits of the reference assumptions.

\takeaway{Imitating normal flows does not ensure normal service interactions. Evasion ought to match client access patterns across services.}

Additional evasion experiments are reported in Appendix \ref{app:robust}.

\subsection{RQ2: Transfer and Boundary Adaptation}
\subsubsection{Transfer without Fine-Tuning}
\label{sec:evo_eval}
Appendix \ref{app:drift} provides more detailed results of the drift experiments.

\paragraph{Transfer without Target Fine-Tuning} We train each model on one dataset and evaluate detection on all other datasets without target-environment fine-tuning. All baseline methods suffer performance degradation, with F1 scores dropping by 55.43 points on average and most falling below 50\% (Figure~\ref{fig:transfer_line_comparison}). In comparison, IDSMorph maintains nearly stable detection performance, with an F1 score degradation of no more than 1 point. IDSMorph keeps its embedding models fixed and constructs dynamic decision boundaries from unlabeled target traffic.

\takeaway{Embedding transfer preserves learned patterns. Target traffic still provide normal references for detection.}

\paragraph{Transfer with Limited Target Labels} We also evaluate adaptation with limited target labels by fine-tuning each baseline model using 5\% of the target domain data. Even with this limited adaptation, all baseline methods still exhibit significant performance degradation, with F1 scores dropping by more than 34.31 points (Figure~\ref{fig:transfer_line_comparison}). In contrast, IDSMorph maintains stable performance without any fine-tuning, using decision boundaries constructed from unlabeled target traffic.

\takeaway{Target labels support adaptation only when they cover normal service accesses. Label quantity cannot resolve missing access patterns.}

\subsubsection{Boundary Adaptation to Network Changes}
\label{sec:temporal}
\paragraph{Periodic Model Adaptation} We evaluate periodic adaptation by fine-tuning baseline models at fixed intervals using data from the current observation window. The limited data within each time window is insufficient to construct a robust representation of traffic behavior (Figure~\ref{fig:transfer_line_comparison}). Consequently, baseline models suffer significant performance drops across most time windows with an overall F1 score decline of 33.13 points. In contrast, IDSMorph updates its decision boundaries without fine-tuning and maintains detection performance across the evaluated windows, with performance degradation of no more than 1 point. The deployment evaluation in Section \ref{sec:realworld} further examines continued detection.

\takeaway{Boundary updates adapt to changing service accesses without requiring the embedding models to relearn flow structures.}

\paragraph{Adaptation to Network Changes} We evaluate adaptation to changes in service grouping and traffic composition by progressively modifying server IP addresses and randomly removing or altering a portion of traffic data. IDSMorph maintains detection performance despite these network changes, with performance degradation of no more than 2 points (Figure~\ref{fig:service_baseline_horizontal}). In contrast, baseline methods exhibit significant performance drops exceeding 34.53 points in the evaluated settings. Appendix \ref{app:drift} reports additional network-condition experiments. Appendix \ref{app:baseline} examines the correspondence between service references and observed services.

\takeaway{Server IP changes need not imply new service behavior. Boundary adaptation should follow access patterns rather than server addresses alone.}

\subsection{RQ3: Embedding and Boundary Contributions}
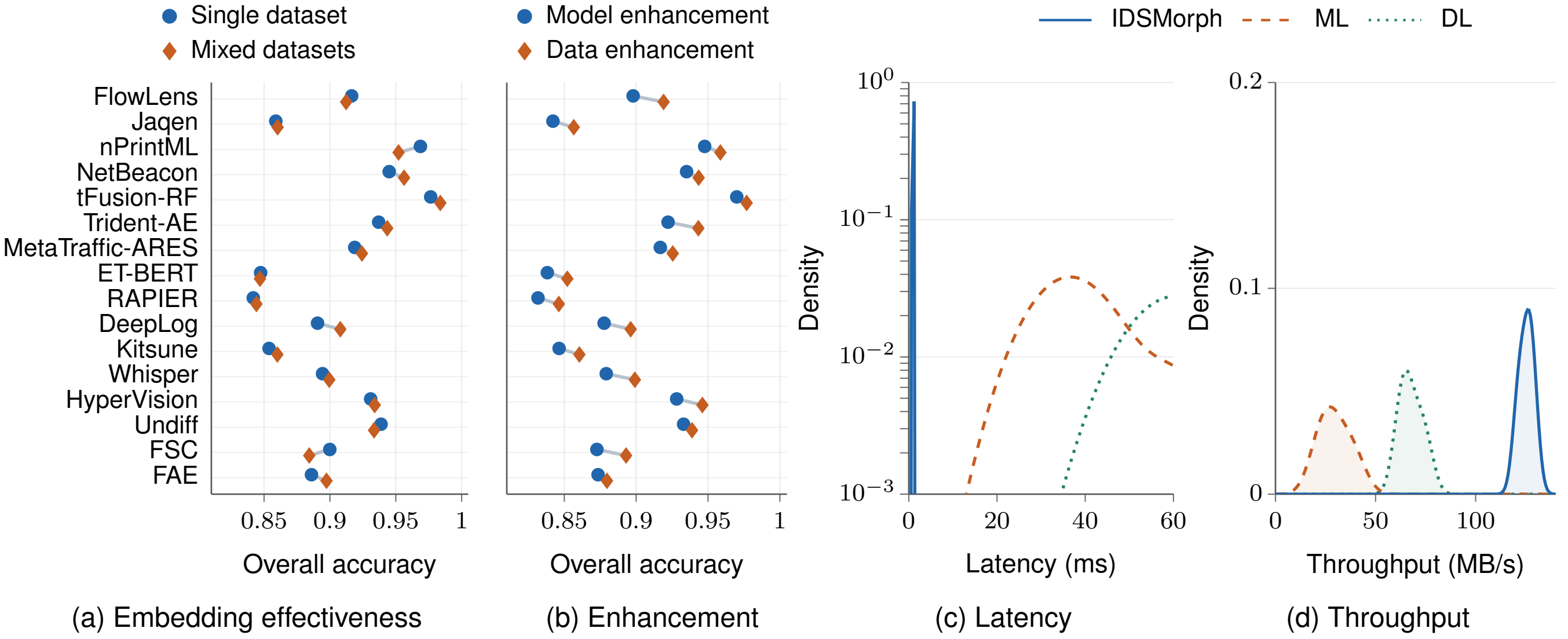
\begin{figure*}[!t]
\centering
\setlength{\evalfullwidth}{\textwidth}
\captionsetup[subfloat]{font=small,farskip=1pt,captionskip=2pt}
\subfloat[Embedding effectiveness\label{fig:accuracy_comparison}]{\makebox[.31\textwidth][c]{\raisebox{0pt}[160pt][34pt]{\begin{tikzpicture}[baseline=(current axis.south)]
\begin{axis}[evaluation categories,typeset ticklabels with strut,width=0.165\evalfullwidth,height=4.7cm,xmin=0.81,xmax=1.005,xtick={0.85,0.9,0.95,1},legend style={at={(0.2,1.025)},anchor=south},legend columns=1,legend cell align=left,column sep=3pt,]
\draw[evalgray,line width=1.1pt] (axis cs:0.9165,-0.12) -- (axis cs:0.9123,0.12);
\draw[evalgray,line width=1.1pt] (axis cs:0.8589,0.88) -- (axis cs:0.8603,1.12);
\draw[evalgray,line width=1.1pt] (axis cs:0.9687,1.88) -- (axis cs:0.9521,2.12);
\draw[evalgray,line width=1.1pt] (axis cs:0.945,2.88) -- (axis cs:0.9563,3.12);
\draw[evalgray,line width=1.1pt] (axis cs:0.9765,3.88) -- (axis cs:0.9838,4.12);
\draw[evalgray,line width=1.1pt] (axis cs:0.937,4.88) -- (axis cs:0.9435,5.12);
\draw[evalgray,line width=1.1pt] (axis cs:0.918836,5.88) -- (axis cs:0.9243,6.12);
\draw[evalgray,line width=1.1pt] (axis cs:0.8474,6.88) -- (axis cs:0.8469,7.12);
\draw[evalgray,line width=1.1pt] (axis cs:0.8418,7.88) -- (axis cs:0.8442,8.12);
\draw[evalgray,line width=1.1pt] (axis cs:0.8906,8.88) -- (axis cs:0.9079,9.12);
\draw[evalgray,line width=1.1pt] (axis cs:0.8538,9.88) -- (axis cs:0.8601,10.12);
\draw[evalgray,line width=1.1pt] (axis cs:0.8944,10.88) -- (axis cs:0.8995,11.12);
\draw[evalgray,line width=1.1pt] (axis cs:0.9309,11.88) -- (axis cs:0.93398,12.12);
\draw[evalgray,line width=1.1pt] (axis cs:0.9388,12.88) -- (axis cs:0.9335,13.12);
\draw[evalgray,line width=1.1pt] (axis cs:0.8999,13.88) -- (axis cs:0.8843,14.12);
\draw[evalgray,line width=1.1pt] (axis cs:0.886,14.88) -- (axis cs:0.8974,15.12);
\addplot[only marks,evalblue,mark=*,mark size=2pt] coordinates {(0.9165,-0.12) (0.8589,0.88) (0.9687,1.88) (0.945,2.88) (0.9765,3.88) (0.937,4.88) (0.918836,5.88) (0.8474,6.88) (0.8418,7.88) (0.8906,8.88) (0.8538,9.88) (0.8944,10.88) (0.9309,11.88) (0.9388,12.88) (0.8999,13.88) (0.886,14.88)};
\addplot[only marks,evalorange,mark=diamond*,mark size=2.4pt] coordinates {(0.9123,0.12) (0.8603,1.12) (0.9521,2.12) (0.9563,3.12) (0.9838,4.12) (0.9435,5.12) (0.9243,6.12) (0.8469,7.12) (0.8442,8.12) (0.9079,9.12) (0.8601,10.12) (0.8995,11.12) (0.93398,12.12) (0.9335,13.12) (0.8843,14.12) (0.8974,15.12)};
\legend{Single dataset,Mixed datasets}
\end{axis}
\pgfresetboundingbox
\path[use as bounding box] ([xshift=-67pt,yshift=-29pt]current axis.south west) rectangle ([xshift=3pt,yshift=34pt]current axis.north east);
\end{tikzpicture}}}}
\hspace{.003\textwidth}%
\subfloat[Enhancement\label{fig:enhancement_dot_comparison}]{\makebox[.205\textwidth][c]{\raisebox{0pt}[160pt][34pt]{\begin{tikzpicture}[baseline=(current axis.south)]
\begin{axis}[evaluation categories,typeset ticklabels with strut,width=0.18\evalfullwidth,height=4.7cm,yticklabels={},xmin=0.81,xmax=1.005,xtick={0.85,0.9,0.95,1},legend style={at={(0.5,1.025)},anchor=south},legend columns=1,legend cell align=left,column sep=3pt,]
\draw[evalgray,line width=1.1pt] (axis cs:0.8979,-0.12) -- (axis cs:0.9191,0.12);
\draw[evalgray,line width=1.1pt] (axis cs:0.8421,0.88) -- (axis cs:0.8566,1.12);
\draw[evalgray,line width=1.1pt] (axis cs:0.9477,1.88) -- (axis cs:0.9586,2.12);
\draw[evalgray,line width=1.1pt] (axis cs:0.9351,2.88) -- (axis cs:0.9435,3.12);
\draw[evalgray,line width=1.1pt] (axis cs:0.9700,3.88) -- (axis cs:0.9769,4.12);
\draw[evalgray,line width=1.1pt] (axis cs:0.9222,4.88) -- (axis cs:0.9433,5.12);
\draw[evalgray,line width=1.1pt] (axis cs:0.9168,5.88) -- (axis cs:0.9255,6.12);
\draw[evalgray,line width=1.1pt] (axis cs:0.8382,6.88) -- (axis cs:0.8521,7.12);
\draw[evalgray,line width=1.1pt] (axis cs:0.8317,7.88) -- (axis cs:0.8462,8.12);
\draw[evalgray,line width=1.1pt] (axis cs:0.8777,8.88) -- (axis cs:0.8962,9.12);
\draw[evalgray,line width=1.1pt] (axis cs:0.8464,9.88) -- (axis cs:0.8605,10.12);
\draw[evalgray,line width=1.1pt] (axis cs:0.8792,10.88) -- (axis cs:0.8991,11.12);
\draw[evalgray,line width=1.1pt] (axis cs:0.9282,11.88) -- (axis cs:0.9461,12.12);
\draw[evalgray,line width=1.1pt] (axis cs:0.9330,12.88) -- (axis cs:0.9389,13.12);
\draw[evalgray,line width=1.1pt] (axis cs:0.8727,13.88) -- (axis cs:0.8930,14.12);
\draw[evalgray,line width=1.1pt] (axis cs:0.8735,14.88) -- (axis cs:0.8797,15.12);
\addplot[only marks,evalblue,mark=*,mark size=2pt] coordinates {(0.8979,-0.12) (0.8421,0.88) (0.9477,1.88) (0.9351,2.88) (0.9700,3.88) (0.9222,4.88) (0.9168,5.88) (0.8382,6.88) (0.8317,7.88) (0.8777,8.88) (0.8464,9.88) (0.8792,10.88) (0.9282,11.88) (0.9330,12.88) (0.8727,13.88) (0.8735,14.88)};
\addplot[only marks,evalorange,mark=diamond*,mark size=2.4pt] coordinates {(0.9191,0.12) (0.8566,1.12) (0.9586,2.12) (0.9435,3.12) (0.9769,4.12) (0.9433,5.12) (0.9255,6.12) (0.8521,7.12) (0.8462,8.12) (0.8962,9.12) (0.8605,10.12) (0.8991,11.12) (0.9461,12.12) (0.9389,13.12) (0.8930,14.12) (0.8797,15.12)};
\legend{Model enhancement,Data enhancement}
\end{axis}
\pgfresetboundingbox
\path[use as bounding box] ([xshift=-3pt,yshift=-29pt]current axis.south west) rectangle ([xshift=3pt,yshift=34pt]current axis.north east);
\end{tikzpicture}}}}
\hspace{.003\textwidth}%
\subfloat[Latency\label{fig:latency_methods}]{\makebox[.24\textwidth][c]{\raisebox{0pt}[160pt][34pt]{\begin{tikzpicture}[baseline=(current axis.south)]
\begin{axis}[evaluation panel,legend columns=3,reverse legend,legend style={at={(1.32,1.10)},anchor=south,column sep=4pt},typeset ticklabels with strut,width=0.17\evalfullwidth,height=4.7cm,xlabel={Latency (ms)},ylabel={Density},xmin=0,xmax=60,ymode=log,ymin=0.001,ymax=1,xtick={0,20,40,60},ytick={0.001,0.01,0.1,1},yticklabels={$10^{-3}$,$10^{-2}$,$10^{-1}$,$10^{0}$},unbounded coords=jump,]
\addplot[evalgreen,dotted,line width=1pt] table[x index=0,y index=1,col sep=comma] {DL_latency.csv};
\addplot[evalorange,dashed,line width=1pt] table[x index=0,y index=1,col sep=comma] {ML_latency.csv};
\addplot[evalblue,solid,line width=1pt] table[x index=0,y index=1,col sep=comma] {DyLine_latency.csv};
\legend{DL,ML,IDSMorph}
\end{axis}
\pgfresetboundingbox
\path[use as bounding box] ([xshift=-31pt,yshift=-29pt]current axis.south west) rectangle ([xshift=3pt,yshift=34pt]current axis.north east);
\end{tikzpicture}}}}
\hspace{.003\textwidth}%
\subfloat[Throughput\label{fig:throughput_methods}]{\makebox[.235\textwidth][c]{\raisebox{0pt}[160pt][34pt]{\begin{tikzpicture}[baseline=(current axis.south)]
\begin{axis}[evaluation panel,typeset ticklabels with strut,width=0.18\evalfullwidth,height=4.7cm,xlabel={Throughput (MB/s)},ylabel={Density},xmin=0,xmax=140,ymin=0,ymax=0.20,xtick={0,50,100},ytick={0,0.1,0.2},]
\addplot[evalgreen,dotted,line width=1pt,fill=evalgreen,fill opacity=0.08] table[x index=0,y index=1,col sep=comma] {DL_throughput.csv};
\addplot[evalorange,dashed,line width=1pt,fill=evalorange,fill opacity=0.08] table[x index=0,y index=1,col sep=comma] {ML_throughput.csv};
\addplot[evalblue,solid,line width=1pt,fill=evalblue,fill opacity=0.08] table[x index=0,y index=1,col sep=comma] {DyLine_throughput.csv};
\end{axis}
\pgfresetboundingbox
\path[use as bounding box] ([xshift=-31pt,yshift=-29pt]current axis.south west) rectangle ([xshift=3pt,yshift=34pt]current axis.north east);
\end{tikzpicture}}}}
\caption{Embedding effectiveness and detection efficiency. (a) Accuracy with IDSMorph embeddings on single and mixed datasets. (b) Accuracy with model and data enhancement, using the method order in (a). (c--d) Latency and throughput distributions for IDSMorph, machine learning (ML), and deep learning (DL); latency density uses a logarithmic scale.}
\label{fig:eff_methods}
\end{figure*}

\paragraph{Embedding Transfer to Other Detectors} IDSMorph's embeddings improve the F1 score of supervised methods by 3.7--4.8 points and that of unsupervised methods by 3.3--6.2 points (Figure~\ref{fig:accuracy_comparison}). On mixed datasets, compared to single-distribution settings, the performance drop for supervised methods is less than 1.6 points, and for unsupervised methods is less than 1.7 points. 

\takeaway{Consistency constraints preserve useful traffic differences across detectors, independently of IDSMorph's boundary construction.}

\paragraph{Representation Enhancement} We compare against: (1) data augmentation schemes, including traffic augmentation \cite{xie2023rosetta,sun2025advtg} and RAPIER's distribution-aware augmentation \cite{qing2023low}; (2) generative model-based approaches that synthesize training samples using GANs \cite{ding2023tmg} and variational autoencoder \cite{zhao2024novel}; and (3) sample quality improvement through integration of the MAWI \cite{mawi} public benign traffic dataset. Although these techniques yield modest improvements, their performance gains consistently fall short of those achieved by IDSMorph embeddings by at least 2.73 points across all datasets and tasks (Figure~\ref{fig:enhancement_dot_comparison}).

\takeaway{Useful traffic augmentation preserves normal access patterns without erasing the differences needed to detect attacks.}

\paragraph{Component Ablation} Removing dynamic decision boundary construction causes the largest F1 decrease, 21.42 points, highlighting its role in adaptation to the deployment environment. Ablating the service interaction consistency builder leads to an F1 decline of 13.31 points. Without the flow structure consistency builder, IDSMorph struggles to construct stable decision boundaries as traffic changes and becomes more sensitive to noise, resulting in an F1 reduction of 6.75 points. We also analyze the four self-supervised augmentation tasks within the flow structure consistency builder and find that omitting any single task degrades performance by 2 to 4 points in F1 score, confirming the complementary value of each augmentation strategy in learning robust behavioral embeddings. Appendix \ref{app:abl} presents detailed ablation studies.

\takeaway{Embeddings distinguish behaviors, while boundaries identify anomalies. Both require comparisons that reflect normal service accesses.}

\subsection{RQ4: Pretraining and Detection Costs}
\label{sec:evo_performance}
We measure the computational and memory efficiency of IDSMorph's model and network components. On a 2.7 GB traffic dataset, IDSMorph has a peak memory usage of 559 MB and processes traffic at an average speed of 145.6 MB/s, with a mean memory consumption of 39 MB during sustained operation. In contrast, the evaluated autoencoder-based baseline requires over 47 GB peak memory and averages 2.5 GB during real-time processing.

\paragraph{Detection Efficiency} We implemented IDSMorph's online detection system using DPDK \cite{dpdk} and compared its inference latency and throughput against baseline methods. IDSMorph processes network traffic at a throughput of 145.6 MB/s, with per-flow decision latency under 1 millisecond (Figures~\ref{fig:latency_methods} and~\ref{fig:throughput_methods}). 

\takeaway{Online detection cost includes adaptation to traffic changes. Boundary updates avoid repeatedly paying the training cost of fixed embedding models.}

\paragraph{Decision Boundary Memory Usage} The decision boundary repository stores compact reference embeddings, keeping memory usage stable as traffic volume increases in the evaluated setting. In baseline experiments on a full day of WIDE backbone network traffic, IDSMorph's peak reference storage remained below 600 MB. As new servers are discovered, the number of active service baselines initially increases, but eventually stabilizes as the set of service IPs converges.

\takeaway{Boundary storage depends on active service groups and retained historical references. Increasing traffic volume need not increase both.}

\paragraph{Pretraining Resource Consumption} Both the flow pattern encoder and the service behavior encoder converge within 30 training epochs. On an NVIDIA GeForce RTX 4090 GPU, the entire pretraining process completes in under 5 minutes, with peak GPU memory usage below 1 GB. Compared with the evaluated pretraining baselines (autoencoders, graph neural networks, and transformers), IDSMorph achieves superior detection performance while consuming less than $0.1\times$ GPU memory.

\takeaway{Pretraining reduces repeated learning only while embeddings remain applicable to observed flow structures and service interactions.}

\section{RQ5: Detection in an Operational Network}
\label{sec:deployment}
\label{sec:realworld}
\subsection{Deployment Setting and Ground Truth}
We evaluate IDSMorph using attack traffic collected within a research organization's network, covering four attack categories: (1) trojan malware attacks \cite{trojansoftware} involving C2 commands and data exfiltration, (2) penetration testing activities \cite{burpsuite} including scanning and reconnaissance, (3) brute-force attacks \cite{scan4all} targeting SSH, FTP, MySQL, and web applications, and (4) DoS attacks \cite{dos} including SYN floods, HTTP floods, and reflection/amplification attacks. The attackers receive the target IP addresses, while the defender has no advance knowledge of attack types or timing. All attacks are precisely labeled by time and IP address and manually verified by the attackers.

The red-team operators independently choose attack strategies and timing against internal test systems in a controlled subnet. Attack times and target IP addresses identify attack traffic for evaluation and are excluded from decision boundary construction. Ethical considerations are provided in Appendix \ref{app:eth}.

\subsection{Detection Results}
The month-long deployment covers 2.268 billion network traffic records involving 247.9K unique IP addresses, including over 352 active client hosts and 105805 server-side IP addresses. Key statistics show that recorded malicious or anomalous traffic accounted for less than 0.0005\% of the total traffic volume.

In real-world deployment, IDSMorph achieved an average precision of 0.9999, a recall of 1.0, an ROC-AUC of 0.9999 and an F1 score of 0.9999. IDSMorph outperformed the evaluated baseline methods.
IDSMorph identified attacks including slow command-and-control (C2) beacons and encrypted data exfiltration.

\takeaway{Encrypted or sparse flows can still deviate from normal service interactions. Low traffic volume alone does not imply benign behavior.}

\subsection{Additional Background-Traffic Changes}
We inject WIDE backbone traffic samples of equal size at different times into the deployment traffic. IDSMorph consistently identifies malicious activities with an F1 score exceeding 99.04\%, demonstrating strong resilience to such drift-induced perturbations. In contrast, baseline methods whose decision logic is tightly coupled to fixed feature distributions suffer severe false positives, with F1 scores dropping below 5\%. Additional deployment results are provided in Appendix \ref{app:real}. 

\takeaway{Rare attacks make false positives critical. Boundary adaptation must distinguish new normal accesses from malicious deviations.}

\section{Discussion and Limitations}
\label{sec:discussion}
\label{app:lim}
\paragraph{Dominant Benign Behavior} IDSMorph requires benign behavior to dominate the local or global groups used as normal references, since each reference is the centroid of the largest cluster. Cross-service comparison can expose a corrupted local reference when other services provide a benign reference, but does not guarantee detection if the attacker also dominates those global groups. Sustained traffic poisoning and coordinated imitation across services remain limitations.

\paragraph{Sparse Services and Cold Start} Sparse observations of less frequently accessed services limit the construction of stable normal references. IDSMorph skips comparisons without an available reference during cold start, which reduces detection coverage until sufficient observations are collected. Global references help only when suitable cross-service observations exist.

\paragraph{Current-Window Observations} Current-window traffic composition affects detection because the same window contributes to both reference construction and flow assessment. Historical boundaries mitigate transient changes, but cannot by themselves ensure a benign reference under sustained attack traffic. Deployment evaluation also depends on ground-truth coverage, since attack records do not establish that all remaining traffic is benign.

\paragraph{Embedding Generalization and Service Variations} Freezing the embedding models avoids repeated training, but does not guarantee generalization to all new protocols or service behaviors. Our boundary analysis also finds lower stability for HTTP/2 traffic, whose multiplexing and stream interleaving increase within-service variation (Appendix \ref{app:baseline}). Improving detection for sparse services and adapting representations to previously unseen communication patterns are directions for future work.

\section{Conclusion}
IDSMorph is a self-adaptive malicious traffic detection system based on consistency-guided dynamic decision boundaries. Flow and interaction representations capture shared communication sequences and access relationships, supporting the construction of local and global decision boundaries for service and client behavior from deployment observations. IDSMorph updates these boundaries without retraining the embedding models. Evaluation on 14 public datasets and a month-long organizational deployment shows the effectiveness of this approach across network environments, traffic changes, and the evaluated evasion strategies. Its effectiveness depends on the availability of representative benign references, particularly for sparse services and groups exposed to sustained malicious traffic.

\bibliographystyle{IEEEtran}
\bibliography{reference}

\appendices
\raggedbottom
\section{Ethical Considerations}
\label{app:eth}
\paragraph{Research Purpose} This study aims to improve malicious traffic detection under changing network conditions. The principal ethical concerns are the privacy of users whose traffic is observed and the potential disruption caused by attack experiments. We address these concerns through controlled attack targets and restrictions on the release of deployment traffic.

\paragraph{Controlled Attack Experiments} The main evaluation uses publicly available datasets. For the organizational deployment, red-team operators independently select attack strategies and timing against internal test systems within a controlled subnet (Section~\ref{sec:realworld}). Attack experiments target only these test systems and exclude publicly accessible devices. This scope allows us to evaluate detection against varied attacks while limiting the systems exposed to the experiments.

\paragraph{Data Privacy} IDSMorph's embedding models use packet sizes, inter-arrival times, directions, and communication relationships without inspecting payload content. These metadata can nevertheless reveal aspects of network activity. We therefore do not release the privacy-sensitive traffic collected during deployment, which is also subject to legal restrictions. Appendix~\ref{app:open} describes the shared artifacts and the data that cannot be made available.

\section{Open Science}
\label{app:open}
\paragraph{Artifacts} We provide source code for IDSMorph and traffic collection, public-data samples, trained models, and configurations for reproducing the experiments.

\paragraph{Data Availability} We evaluate IDSMorph using public datasets and provide sample excerpts. Deployment traffic cannot be shared because it contains privacy-sensitive information and is subject to legal restrictions.

\paragraph{Trained Models} Model checkpoints and configuration files are included in the repository.

\section{Parameter Settings and Sensitivity Analysis}
\label{sec:appendix1}
\subsection{Default Parameter Settings}
\begin{list}{\arabic{enumi})}{%
  \usecounter{enumi}%
  \setlength{\leftmargin}{0pt}%
  \setlength{\rightmargin}{0pt}%
  \setlength{\labelwidth}{1.1em}%
  \setlength{\labelsep}{0.3em}%
  \setlength{\itemindent}{1.4em}%
  \setlength{\listparindent}{0pt}%
  \setlength{\topsep}{2pt}%
  \setlength{\partopsep}{0pt}%
  \setlength{\itemsep}{2pt}%
  \setlength{\parsep}{0pt}%
}
\item \textbf{Flow Structure Consistency Builder.} We train for 100 epochs using AdamW with a learning rate of $10^{-5}$.
\item \textbf{Service Interaction Consistency Builder.} The architecture uses a CNN-to-linear embedding pipeline and a graph attention layer (GAT) with 8 heads.
\item \textbf{Similarity and Clustering.} The detection contrast similarity threshold is 0.95. DBSCAN uses a neighborhood radius of 0.5 and a minimum of 5 samples.
\end{list}

\subsection{Parameter Sensitivity Analysis}
\paragraph{Flow Similarity Threshold} We evaluate how the flow similarity threshold affects service grouping and detection performance (Table~\ref{table_hyper_thre}).

\renewcommand{\arraystretch}{1.25}
\newcommand{\tablefont}{\small\bfseries}
\newcolumntype{d}{S[table-format=1.4]}
\newcolumntype{D}{S[table-format=3.0]}
\newcommand{\bestrow}{\rowcolor{yellow!25}}
\captionsetup[table]{position=top,skip=6pt,font=small,labelfont=bf}

\begin{table}[!t]
  \centering
  \footnotesize
  \setlength{\tabcolsep}{3pt}
  \caption{Effect of the flow similarity threshold on detection and service grouping. ACC, Pre, and REC denote accuracy, precision, and recall. The highlighted row has the highest F1 score among the evaluated settings.}
  \label{table_hyper_thre}
  \begin{tabular*}{\columnwidth}{@{\extracolsep{\fill}}lddddD@{}}
    \toprule
    {\textbf{Thresholds}} & {\textbf{ACC}} & {\textbf{Pre}} & {\textbf{REC}} & {\textbf{F1}} & {\makecell{\textbf{Number of}\\\textbf{service groups}}} \\
    \midrule
    0.6  & 0.7254 & 0.7157 & 0.7525 & 0.7336 & 154 \\
    0.7  & 0.8039 & 0.7771 & 0.8514 & 0.8125 & 285 \\
    0.8  & 0.9056 & 0.8732 & 0.9504 & 0.9102 & 320 \\
    \bestrow
    0.9  & 0.9999 & 0.9997 & 0.9994 & 0.9994 & 377 \\
    0.95 & 0.9594 & 0.9459 & 0.9745 & 0.9601 & 532 \\
    \bottomrule
  \end{tabular*}
\end{table}

\noindent Lower thresholds merge unrelated flows, whereas higher thresholds split correlated flows. Increasing the threshold reduces traffic relations and increases the number of service groups. Performance peaks at 0.9; higher thresholds produce excessive isolated flows, making contrastive learning less stable and more susceptible to noise.

\takeaway{Service grouping defines normal comparisons. Boundary construction cannot repair groups that merge unrelated service accesses.}

\begin{figure*}[!t]
\centering
\setlength{\evalfullwidth}{\textwidth}
\captionsetup[subfloat]{font=small,farskip=1pt,captionskip=2pt}
\makebox[\textwidth][l]{\fontsize{8}{9}\selectfont\sffamily\textcolor{evalblue}{\rule{10pt}{1pt}}~Accuracy\quad\textcolor{evalorange}{\rule{10pt}{1pt}}~Precision\quad\textcolor{evalgreen}{\rule{10pt}{1pt}}~Recall\quad\textcolor{evalpurple}{\rule{10pt}{1pt}}~F1}\par\vspace{2pt}
\subfloat[History size\label{fig:hyperparam_history}]{\makebox[.247\textwidth][c]{\raisebox{0pt}[98pt][29pt]{\begin{tikzpicture}[baseline=(current axis.south)]
\begin{axis}[evaluation panel,width=0.185\evalfullwidth,height=3.2cm,xlabel={History size $K$},ylabel={Score},xmode=log,log basis x=2,xmin=1.8,xmax=18,xtick={2,4,8,16},xticklabels={2,4,8,16},ymin=0.8,ymax=1.01,ytick={0.8,0.9,1},]
\addplot[evalblue,mark=*,line width=0.85pt,mark size=1.8pt] coordinates {(2,0.8501) (4,0.9044) (8,0.9999) (16,0.9706)};
\addplot[evalorange,mark=square*,dashed,line width=0.85pt,mark size=1.8pt] coordinates {(2,0.8187) (4,0.8667) (8,0.9998) (16,0.9662)};
\addplot[evalgreen,mark=triangle*,densely dotted,line width=0.85pt,mark size=1.8pt] coordinates {(2,0.9001) (4,0.9557) (8,0.9994) (16,0.9754)};
\addplot[evalpurple,mark=diamond*,dashdotted,line width=0.85pt,mark size=1.8pt] coordinates {(2,0.8578) (4,0.9091) (8,0.9994) (16,0.9707)};
\end{axis}
\end{tikzpicture}}}}
\hspace{.004\textwidth}%
\subfloat[Detection threshold\label{fig:hyperparam_threshold}]{\makebox[.247\textwidth][c]{\raisebox{0pt}[98pt][29pt]{\begin{tikzpicture}[baseline=(current axis.south)]
\begin{axis}[evaluation panel,width=0.185\evalfullwidth,height=3.2cm,xlabel={Threshold},ylabel={Score},xmin=0.68,xmax=1.02,xtick={0.7,0.8,0.9,1},ymin=0.8,ymax=1.01,ytick={0.8,0.9,1},]
\addplot[evalblue,mark=*,line width=0.85pt,mark size=1.8pt] coordinates {(0.7,0.9061) (0.8,0.9999) (0.9,0.9901) (1.0,0.9857)};
\addplot[evalorange,mark=square*,dashed,line width=0.85pt,mark size=1.8pt] coordinates {(0.7,0.9101) (0.8,0.9901) (0.9,0.9959) (1.0,0.9904)};
\addplot[evalgreen,mark=triangle*,densely dotted,line width=0.85pt,mark size=1.8pt] coordinates {(0.7,0.9009) (0.8,0.9999) (0.9,0.9859) (1.0,0.9801)};
\addplot[evalpurple,mark=diamond*,dashdotted,line width=0.85pt,mark size=1.8pt] coordinates {(0.7,0.9059) (0.8,0.9994) (0.9,0.9908) (1.0,0.9856)};
\end{axis}
\end{tikzpicture}}}}
\hspace{.004\textwidth}%
\subfloat[IDSMorph memory\label{fig_usg_Chameleon}]{\makebox[.247\textwidth][c]{\raisebox{0pt}[98pt][29pt]{\begin{tikzpicture}[baseline=(current axis.south)]
\begin{axis}[evaluation panel,width=0.185\evalfullwidth,height=3.2cm,xlabel={Time (h)},ylabel={Memory (MB)},xmin=0,xmax=26.67,xtick={0,10,20},ymin=0,ymax=800,ytick={0,400,800},]
\addplot[evalblue,line width=1pt,mark=*,mark size=1.3pt,mark repeat=4] coordinates {(0,0) (1,100) (2,200) (3,300) (4,400) (5,450) (6,500) (7,550) (8,600) (9,620) (10,650) (11,680) (12,700) (13,720) (14,740) (15,750) (16,760) (17,770) (18,780) (19,750) (20,700)};
\end{axis}
\end{tikzpicture}}}}
\hspace{.004\textwidth}%
\subfloat[DL baseline memory\label{fig_usg_deepl}]{\makebox[.247\textwidth][c]{\raisebox{0pt}[98pt][29pt]{\begin{tikzpicture}[baseline=(current axis.south)]
\begin{axis}[evaluation panel,width=0.185\evalfullwidth,height=3.2cm,xlabel={Time (h)},ylabel={Memory (GB)},xmin=0,xmax=26.67,xtick={0,10,20},ymin=0,ymax=50,ytick={0,25,50},x filter/.code={\pgfmathparse{\pgfmathresult/60}},y filter/.code={\pgfmathparse{\pgfmathresult/1000}},]
\addplot[evalorange,line width=1pt,mark=square*,mark size=1.3pt,mark repeat=5] coordinates {(0,0) (50,10000) (100,30000) (150,35000) (200,37000) (250,38000) (300,38000) (350,38000) (400,38000) (450,38000) (500,38000) (550,38000) (600,38000) (650,38000) (700,38000) (750,38000) (800,38000) (850,38000) (900,38000) (950,38000) (1000,38000) (1050,38000) (1100,38000) (1150,38000) (1200,38000) (1250,38000) (1300,38000) (1350,38000) (1400,38000) (1450,38000) (1500,38000)};
\end{axis}
\end{tikzpicture}}}}
\caption{Parameter sensitivity and deployment memory. (a--b) Detection scores across evaluated history sizes and thresholds; history size uses a base-2 logarithmic scale. (c--d) Memory usage over the same time range, measured in MB for IDSMorph and GB for the deep learning baseline.}
\label{fig:hyperparam_analysis}\label{fig:mem_comparison}
\end{figure*}
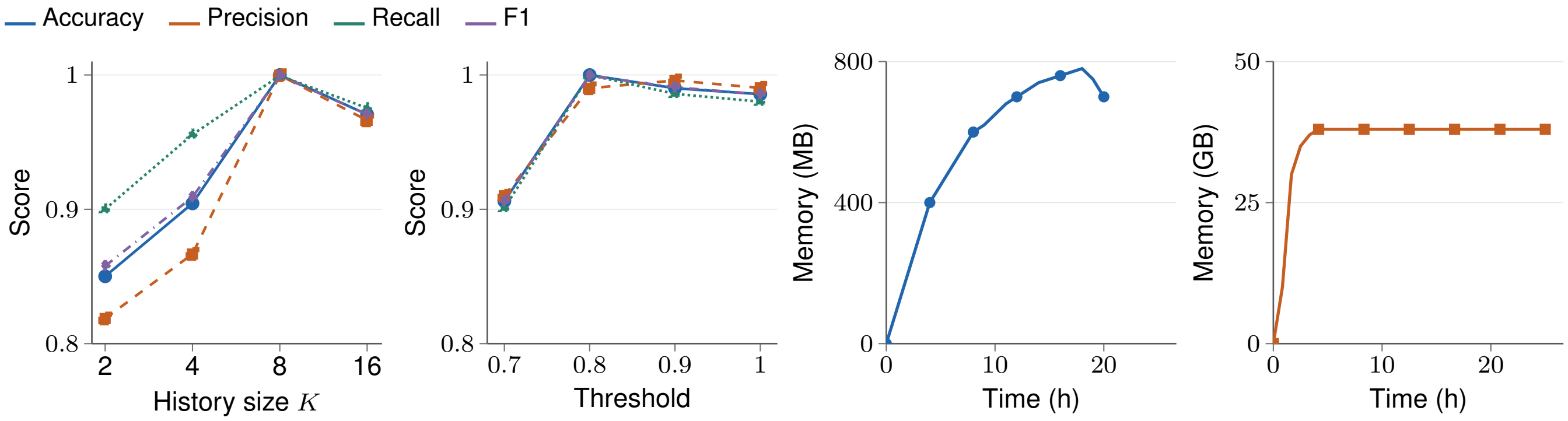

\paragraph{Boundary History Size} The number of retained historical references determines how boundary updates balance transient noise against changes in normal behavior (Figure~\ref{fig:hyperparam_analysis}). Too few references make the boundaries susceptible to noise, whereas too many retain outdated patterns and increase false positives as normal services change. The best performance is achieved with a history size of 8.

\takeaway{Boundary history should reflect how long normal access patterns persist, rather than how much traffic arrives.}

\paragraph{Detection Threshold} The detection threshold controls the deviation from a normal reference required to flag a flow. We evaluate classification performance across threshold values (Figure~\ref{fig:hyperparam_analysis}). The model achieves optimal detection performance when the confidence threshold is set to 0.8. Owing to the precisely constructed dynamic baselines of IDSMorph, the variation in threshold values remains constrained within a narrow distribution range.

\takeaway{Normal behavior centers and detection thresholds serve different roles. Updating a center does not determine the acceptable deviation from it.}

\subsection{Data and Model Augmentation Configuration}
\paragraph{Data Augmentation} Hybrid augmentation combines Rosetta and AdvTG, each generating 50\% of the augmented samples to diversify traffic semantics while preserving protocol validity. RAPIER synthesizes samples according to each service class's data distribution to improve boundary generalization.

\paragraph{Model-Based Augmentation} We train a GAN \cite{ding2023tmg} and a variational autoencoder \cite{zhao2024novel} on the original traffic data to generate benign flows.

\paragraph{Additional Benign Traffic} We cluster MAWI traffic \cite{mawi} and add semantically similar benign flows to training, broadening coverage while preserving service-level consistency.

\section{Ablation Analysis}
\subsection{Ablation Study on Model Components and Design Choices}
\label{app:abl}
The ablation results show that embedding architectures, training objectives, sampling, and boundary construction each affect detection performance. Replacing the embedding module with alternative architectures (VAE, Transformer, and TrafficFormer) reduces performance by more than 5.24 points. When replacing the multi-objective loss function of the flow structure consistency builder with MSE, the model is still able to learn reasonable embedding representations but struggles to discriminate between different embedding relationships, resulting in a drop of 3.87 points in detection performance. Replacing IDSMorph's service interaction consistency builder with graph clustering or graph autoencoder approaches leads to a performance drop of 14.82 points, because these alternatives are more susceptible to noise. Replacing IDSMorph's sampling strategy with address-based contrastive learning \cite{fu2025training} results in a performance drop of 15.22 points, confirming the advantage of aligning intra-service and inter-service behaviors and patterns through IDSMorph's designed contrastive framework. Finally, replacing IDSMorph's multiple dynamic baselines with a single baseline derived from embedding clustering leads to a performance drop of 29.85 points, as such coarse-grained baselines fail to capture the multi-modal patterns of normal traffic behavior. 
These substitutions probe different failure modes. Architecture and loss ablations test the learned representations; sampling tests whether comparisons extend beyond server identity. The single-baseline ablation tests comparison scope after embedding training. Its larger degradation indicates that representation quality alone is insufficient: collapsing distinct normal access patterns can obscure the deviations used for detection.

\takeaway{Distinct service accesses need appropriate comparisons. Better embeddings cannot recover deviations hidden by a shared boundary.}

\section{Additional Transfer and Network-Change Results}
\label{app:drift}
\subsection{Target-Data Availability}
Baseline adaptation remains sensitive to target-distribution differences even when labeled target data is available. We vary the proportion of target data used for fine-tuning in the transfer experiments. First, when the source and target datasets share attack instances of the same category, some models achieve a performance gain of over 5.6 points. However, other models exhibit an F1 score decrease of 27.19 points when the distribution shift exceeds 27.85\%, indicating overfitting to specific attack instances rather than generalizing to the underlying attack semantics. Across multiple overlapping data ratios, baseline methods consistently trigger high false positive rates exceeding 45.74\%. This stems from their reliance on static, closed-world assumptions that define benign behavior through fixed decision boundaries, leading to erratic and unpredictable error patterns under distribution shifts. The results show the sensitivity of these detectors to changes in the target distribution. In contrast, IDSMorph's behavior-based dynamic baselines remain stable under such drift conditions. IDSMorph updates its decision boundaries without retraining or labeled target data.

\takeaway{Shared attack categories do not ensure transfer across networks. Detection also depends on how target traffic defines normal service accesses.}

\subsection{Observation Window Size}
Observation window size affects both responsiveness to traffic changes and the data available for model adaptation. Larger windows lead to significantly degraded performance with accuracy drops exceeding 47.82 points, since they average over rapid traffic shifts and fail to capture short-term environmental changes. Conversely, using very small windows and frequently retraining models incurs prohibitive computational overhead and, more critically, suffers from severe performance degradation (over 50.84 points) due to the scarcity of labeled data in each narrow window, which prevents stable model convergence.

\takeaway{Frequent boundary updates require sufficient normal accesses. Short windows can limit adaptation by reducing available observations.}

\subsection{Impact of Complex Network Conditions}
IDSMorph retains detection performance under the evaluated transmission perturbations. We replay traffic with MTU-induced packet fragmentation, network congestion, and packet reordering to test changes between data collection and deployment conditions. Under these perturbations, all baseline methods suffer severe performance degradation, with F1 scores dropping by more than 74.58 points. This decline stems from their reliance on fixed and sensitivity-prone feature representations. In contrast, IDSMorph exhibits minimal performance loss, no more than 1 point, owing to its focus on service-level interaction semantics, which enables it to maintain robustness even in complex and noisy network conditions.

\takeaway{Packet sequence changes can reflect transmission perturbations rather than attacks. Detection must preserve this distinction.}

\section{Robustness Analysis}
\label{app:robust}
\subsection{Additional Evasion Attacks}
\paragraph{Packet-Level Imitation} The packet-level imitation attack uses full knowledge of benign traffic patterns to craft malicious flows that resemble normal traffic. Experimental results show that even under such strong mimicry attacks, IDSMorph's detection performance degrades by less than 2 points under the evaluated packet-level imitation attack.

Packet-level resemblance can leave differences in service access sequences and relationships. The result supports complementary interaction evidence, rather than establishing that the flow representation alone resists this attack.

\takeaway{Matching flow structure can leave service interactions unmatched. Evasion depends on which access relationships remain different.}

\paragraph{Interaction-Level Imitation} The interaction-level imitation attack reproduces both traffic content and observed legitimate user interaction patterns. Even under this challenging setting, IDSMorph's detection performance degrades by less than 3 points. This result concerns the implemented imitation strategy and does not establish robustness against all interaction-level attacks.

Under the tested strategy, simulated communications still differ from legitimate multi-step interactions. Coordinated imitation that also matches cross-service client behavior may remove this evidence and remains a limitation (Section~\ref{sec:discussion}).

\takeaway{Local and global boundaries require behavioral differences. Imitating all observed service accesses can remove detection evidence.}

\subsection{Stability of Embedding Models}
IDSMorph's detection performance remains stable across the evaluated pretraining datasets. Self-supervised augmentation guides the models to learn shared sequence and interaction patterns. We train IDSMorph on multiple WIDE backbone network traffic datasets, and the resulting detection performance varies by less than 1 point across them, highlighting its strong stability and generalization capability. In contrast, baseline methods \cite{lin2022bert,mirsky2018kitsune} require dataset-specific fine-tuning to maintain performance, as their effectiveness degrades significantly when applied across different network environments without re-adaptation.

This experiment concerns encoder training data, whereas target-network adaptation concerns the references constructed during deployment. Similar performance across WIDE traces supports feature reuse, but does not guarantee coverage of unseen protocols or sufficient observations for every service.

\takeaway{Stable source embeddings do not guarantee detection of deviations in unseen flow structures or service interactions.}

\section{Additional Deployment Results}
\label{app:real}
\subsection{Resource Consumption Analysis}
IDSMorph maintains stable memory usage during deployment as additional service references are constructed. Initial deployment produced 37687 dynamic, service-specific references. Memory measurements demonstrate that IDSMorph maintains highly efficient resource utilization (Figures~\ref{fig_usg_Chameleon} and \ref{fig_usg_deepl}). IDSMorph achieves the desired detection performance with a peak memory footprint of 745 MB, and maintains an average memory usage of 39 MB during real-time traffic processing at a rate of 145.6 MB/s. By comparison, the AE-based method \cite{mirsky2018kitsune} requires significantly more resources, peaking at 47132 MB under ideal conditions. When processing equivalent real-time traffic, its average memory consumption escalates to 2482 MB, which presents considerable challenges for complex, large-scale network detection scenarios.

\takeaway{Traffic volume drives processing demand, while service groups and boundary history drive reference storage. Deployment must account for both.}

\section{Decision Boundary Analysis}
\label{app:baseline}
The boundary analysis checks whether normal references correspond to the observed network services. We query reference IP addresses using Nslookup and extract the Server Name Indication field from TLS handshake data. For traffic lacking an SNI field, we manually access the corresponding network services and compare the observed traffic with the constructed baseline. The results show that 98.44\% of the labeled services match their respective baselines precisely, confirming the semantic fidelity and practical accuracy of IDSMorph's service modeling. References for HTTP/2 traffic exhibit lower stability because frame multiplexing and variable stream interleaving increase variation among accesses to the same service.

Service matching checks reference identity, whereas the HTTP/2 observation concerns variation among accesses. The 98.44\% match rate therefore measures semantic agreement, not detection accuracy or boundary stability. Multiplexed services may require several representative access patterns even when their identity is known.

\takeaway{One service can support multiple normal access patterns. Correct service grouping alone does not ensure a stable decision boundary.}

\section{Additional Traffic Classification Experiment}
\label{app:classification}
We also evaluate the flow and interaction representations in an additional encrypted traffic classification task. We extend IDSMorph's framework by constructing per-application single-class embedding baselines and assign flows to the nearest application cluster. In experiments on Android \cite{mahdavifar2020dynamic} and VPN \cite{gil2016characterization} traffic classification, this approach achieves an F1 score 3 points higher than MetaTraffic, a SOTA method for encrypted traffic recognition. 

The classification experiment changes the reference granularity from normal service behavior to application identity. Assigning a flow to the nearest application cluster asks which known application it resembles, whereas malicious traffic detection asks whether it deviates from available normal references. The improvement supports reuse of the representations across these tasks, without making application recognition equivalent to attack detection.

\takeaway{Application references support traffic classification. Detecting attacks requires normal service and client references instead.}

\end{document}